\documentclass[10pt,journal,compsoc]{IEEEtran}

\usepackage{cite}
\usepackage{amsmath,amssymb,amsfonts,upgreek,diagbox}
\usepackage{algorithm}
\usepackage{array}

\usepackage{algorithmic}
\usepackage{cite}
\usepackage{graphicx}
\usepackage{textcomp,makecell}
\usepackage{xcolor}
\usepackage{diagbox}
\usepackage{subfigure}
\usepackage{epstopdf}
\usepackage[justification=centering]{caption}
\usepackage{multirow}
\usepackage{makecell}
\usepackage{enumerate}
\usepackage{url}
\usepackage{color}
\usepackage{diagbox}
\usepackage{CJK}
\usepackage{amsthm}

\pdfoutput=1

\begin{document}
	
\title{Privacy-Preserving Image Retrieval Based on Additive Secret Sharing}

\author{Zhihua Xia,~\IEEEmembership{Member,IEEE}, Qi Gu, Lizhi Xiong, Wenhao Zhou, Jian Weng,~\IEEEmembership{Member,IEEE}

\thanks{Zhihua Xia (corresponding author, email: xia$\_$zhihua@163.com), Qi Gu (email: guqi634337549@163.com), Lizhi Xiong and Wenhao Zhou are with Engineering Research Center of Digital Forensics, Ministry of Education, School of Computer and Software, Jiangsu Engineering Center of Network Monitoring, Jiangsu Collaborative Innovation Center on Atmospheric Environment and Equipment Technology, Nanjing University of Information Science \& Technology, Nanjing, 210044, China}
\thanks{Jian Weng is with Jinan University, Guangzhou 510632, China}}

\markboth{Privacy-Preserving Image Retrieval Based on Additive Secret Sharing}%
{Shell \MakeLowercase{\textit{et al.}}: Bare Demo of IEEEtran.cls for IEEE Journals}
	
\IEEEtitleabstractindextext{
\begin{abstract}
	
The rapid growth of digital images motivates individuals and organizations to upload their images to the cloud server. To preserve the privacy, image owners would prefer to encrypt the images before uploading, but it would strongly limit the efficient usage of images. Plenty of existing schemes on privacy-preserving Content-Based Image Retrieval (PPCBIR) try to seek the balance between security and retrieval ability. However, compared to the advanced technologies in CBIR like Convolutional Neural Network (CNN), the existing PPCBIR schemes are far deficient in both accuracy and efficiency. With more cloud service providers, the collaborative secure image retrieval service provided by multiple cloud servers becomes possible. In this paper, inspired by additive secret sharing technology, we propose a series of additive secure computing protocols on numbers and matrices with better efficiency, and then show their application in PPCBIR. Specifically, we extract CNN features, decrease the dimension of features and build the index securely with the help of our protocols, which include the full process of image retrieval in plaintext domain. The experiments and security analysis demonstrate the efficiency, accuracy and security of our scheme.

\end{abstract}
\begin{IEEEkeywords}
Privacy-preserving Image Retrieval, Additive Secret Sharing, Pre-trained CNN, Secure PCA, Secure Index Building
\end{IEEEkeywords}
}

\maketitle

\section{Introduction}\label{sec:Introduction}

\IEEEPARstart{R}{ecent} years have witnessed the explosive growth of personal image, and services like CBIR are utilized by more and more people with the help of cloud computing. Accordingly, the CBIR technologies attract plenty of researchers. The schemes on feature extraction \cite{zheng2017sift} from images and index building \cite{fu2019satellite} for large-scale vectors have been researched in detail and these excellent schemes have been put into practice like Google search by image.

It should be noticed that images contain rich sensitive information in many cases. For instance, patients would not like to expose their medical images. As the Cloud Server (CS) is always not fully trusty, it is unsuitable to upload the plaintext images to CS directly. Therefore, more and more researchers pay their attention to PPCBIR. The PPCBIR schemes should not only keep the accuracy and efficiency of image retrieval, but also ensure the images and their features will not be exposed under the reasonable security assumption.

However, due to the security restrictions, the accuracy and efficiency of existing PPCBIR schemes are much lower than works \cite{wei2017selective} in CBIR. For instance, the features used in PPCBIR either put a heavy burden \cite{lu2009enabling} on the image owner or show inferiority \cite{xia2019boew} during the retrieval. The main reason for the inferiority is the plenty of non-linear computation appeared in the state-of-the-art CBIR schemes. It should be noticed that the traditional functional encryption tools are quite low efficient when facing these computations.

With more and more different cloud server manufacturers are willing to provide cloud computing services, the service provided by two non-collusion CS attracts many researchers in recent years. The attendance of two non-collusion CS can avoid the utilization of traditional high-complexity functional encryption methods, which gives a new way to decrease the consumption of secure retrieval.

Additive secret sharing is a typical technology which widely used in Secure Multi-party Computation (SMC). Recently, more researchers utilize this technology to execute their tasks \cite{huang2019lightweight,liu2019toward,chen2020shosvd} under the environment of cloud computing. Especially, recent work \cite{xiong2020efficient} constructs protocols that can support the secure computation on almost all basic elementary functions in constant-rounds interaction based on additive secret sharing. However, the existing schemes always ignore secure matrix computation and cause unnecessary communication complexity for ensuring the information-theoretic security of input.

Accordingly, in this work, we aim to bridge the gap between PPCBIR and CBIR based on additive secret sharing technology. We first construct some basic secure computation protocols based on two non-collusion servers, then show how to utilize them in PPCBIR. In summary, we mainly make the following contributions:

\begin{itemize}

\item[1)] We construct a series of novel protocols based on additive secret sharing. Especially, besides a more efficient secure comparison method, the computation protocol on matrices like matrix inversion is also designed to support complex computation.

\item[2)] Secure CNN features are extracted as visual descriptors for image retrieval from a pre-trained VGG16 by using the proposed protocols. Due to the efficiency of our protocols, our scheme consumes much less time than a recent CNN feature extraction method\cite{huang2019lightweight}.

\item[3)] Based on the secure matrix inversion protocol, we further propose the scheme which could execute secure feature compression, e.g., Principle Component Analysis (PCA), within constant rounds interaction. We also expand the comparison protocol to secure sorting for better supporting index building on encrypted features. To our knowledge, this is the first work that considers the way of compressing encrypted features in PPCBIR. The utilization of compression and index greatly decrease the consumption during the secure retrieval.

\item[4)] We test the proposed scheme on two real-world image databases and demonstrate that our scheme is far better than the previous PPCBIR schemes in both accuracy and efficiency. Especially, the accuracy is consistent with the plaintext state, which verifies the lossless property of our protocols.

\end{itemize}

The rest of our paper is organized as follows. In section \ref{sec:related work}, we briefly introduce the existing schemes on CBIR and PPCBIR. Some preliminaries will be described in section \ref{sec:Prelimin}. The system model will be given in section \ref{sec:Sysmodel}. Some basic protocols will be proposed in section \ref{sec:secretMatrixComputation}. Section \ref{sec:featurextrac} gives the details of our scheme. The security analysis and experiment results will be shown in section \ref{sec:SecuAnal} and \ref{sec:experimentRes}. Finally, we make the conclusion in section \ref{sec:conclu}.

\section{Related work}\label{sec:related work}

We introduce the existing work related to this paper from the perspective of CBIR and PPCBIR.

\subsection{Content-Based Image Retrieval}\label{subsec:CBIR}
In the early stage of CBIR, the global features \cite{jain1996image}, such as color and texture, are firstly extracted by researchers. However, as the global feature is fragile to the influence like illumination or rotation. Benefiting from the scale-space theory, local features\cite{lowe1999object} (e.g., SIFT) are proposed and show a great advantage to global features. Some following local features like SURF \cite{bay2006surf} try to accelerate the extraction process. However, local features extracted from images are large and unstable, in this case, the actual retrieval time consumption will be unacceptable. Many feature aggregation methods, like Bag-Of-Words (BOW)\cite{sivic2003video}, are proposed to cope with this problem. During the \emph{offline phase} (i.e., before retrieval), a codebook is trained and the local features of each image will be finally encoded as the unified feature vector.

With the rapid development of CNN, the feature extracted by the pre-trained CNN shows a great advantage\cite{uijlings2013selective} to traditional hand-crafted features. Although the neural network is widely used in plenty of supervised learning tasks, as CBIR is typically an unsupervised task, pre-trained CNN is more reasonable for the image retrieval scene. Interestingly, Babenko \textit{et al.} \cite{babenko2014neural} found that the CNN trained in a quite different large image database (e.g., ImageNet dataset\cite{russakovsky2015imagenet}) can still show excellent retrieval accuracy in a new dataset. In the later research, Uijlings \textit{et al.} \cite{uijlings2013selective} noticed that the feature extracted from the convolutional layer is better than the final fully connected layer. However, similar to the local feature, the feature vectors extracted from the convolutional layer are in high dimension. For instance, the size of features, extracted from the last convolutional layer in VGG16, will be $512 \times M/16 \times N/16$, where $M$, $N$ is the length and width of the input color image. Babenko \textit{et al.} \cite{babenko2015aggregating} further noticed that the simpler aggregation method, like computing the mean value of each dimension as the feature, is more suitable for deep features. The above deep descriptor can be aggregated as a 512-dim (dimension) feature for each image. Although some later works\cite{wei2017selective} propose more optimization tricks, the improvement of accuracy is no longer outstanding for common images.

A problem closely related to CBIR is Nearest-Neighbor Search (NNS). After getting image features, it is still a difficult problem to search for some nearest images from a database that contains large-scale images efficiently. Accurate search is always time-consuming due to the "curse of dimension", and the mainstream schemes on NNS try to cope with the Approximate NNS (ANNS) problem. To our knowledge, the schemes in ANNS can be classified into the four categories: tree-based index\cite{muja2014scalable}, hash-based index\cite{gan2012locality}, Product Quantization (PQ\cite{jegou2010product}), and graph-based index\cite{fu2019satellite}.

The kernel task in ANNS is reducing the computational complexity of retrieval by constructing an index in the \emph{offline phase}. Tree-based schemes achieve the goal by splitting the original space into tree structure subspaces. During the retrieval, only points in partial subspaces will be involved in the actual distance computation. Therefore the most significant problem is how to divide the original space. One typical scheme called HKM (Hierarchical $k$-Means\cite{nister2006scalable,muja2014scalable}) tries to use typical $k$-means to construct the index. This scheme firstly uses $k$-means to divide all the data into $k$ different categories. Then, for each category, if the vectors in the current category are too many, the server will use $k$-means to split this category again. The index building process will be executed until the number of vectors in each category less than a fixed number.

Hash-based schemes follow the idea that the neighbor points in high dimension will still be neighbors when they are mapped into the low dimension. They build a reasonable mapping method called Locality Sensitive Hash (LSH) family\cite{datar2004locality} for common distance measurements (e.g., Euclidean distance). The original vector will be mapped to an integer number, and the similar images have a higher probability to collide (i.e., same hash value) with each other. One typical LSH method called C2LSH (Count Collusion LSH\cite{gan2012locality}) proposed the concept of dynamic collision and virtual rehash, which got significant storage and accuracy advantage. Briefly speaking, they construct a series of LSH mapping functions in the same LSH family, then use these functions to map all vectors in the database. The vectors which have higher collision frequency will have the chance to participate in actual distance computation. We will describe more calculation details of HKM and C2LSH in subsection \ref{subsec:HKMAndC2LSH}.

In recent years, great progress has been made in the PQ-based and graph-based schemes, which make billion-scale vector retrieval practical. Although the motivation behind these methods is far different from the former two categories, however, the actual calculation is similar (e.g., multiplication, comparison, etc.). For simplicity and generality, in this paper, we choose two typical methods HKM and C2LSH as examples and apply them in encrypted image feature vectors index building. It will be easy to notice that other ANNS schemes are also able to incorporate into our work.

\subsection{Privacy-preserving Content-Based Image Retrieval}\label{subsec:PPCBIR}

For security considerations, the PPCBIR task has attracted more and more researchers in this decade. Different from CBIR, the schemes in PPCBIR need to protect the image and its feature for the owner of the image and provide the CBIR service to authorized users with the help of the cloud server.

Existing schemes on PPCBIR can be classified into two categories: \emph{feature-encryption based schemes} and \emph{image-encryption based schemes}. In the first category\cite{lu2009enabling,xia2015towards,yuan2015seisa}, the image owner needs to extract the features from plaintext images and builds an index for them. Then the owner uploads encrypted images and a corresponding index to the cloud server. Since the feature extraction and index building are always resource-consuming tasks, it is not appropriate for leaving them to the image owner.

In this case, schemes\cite{xia2019boew,cheng2016encrypted} in the second category, which outsources these tasks, attract more researchers in recent years. In this category, the image owner only needs to encrypt his images and outsources them to CS. The key problem here is how to extract the valid feature from an encrypted image. In this case, schemes in the second category can be classified into two classes based on the type of encrypted features that CS extracts from the encrypted images. The detailed information is described as follows.

\subsubsection{schemes based on statistic feature}

The schemes in the first class extract statistic histograms as the encrypted image feature. The main method here is protecting the images and their features by permutation and keeping valid distance between encrypted images. Ferreira \textit{et al.} \cite{ferreira2017practical} proposed a tailor-made encryption scheme for images. In this scheme, the image color is protected by value substitution and image texture is shuffled randomly by rows and columns. After uploading, the HSV (Hue-Saturation-Value) color histograms are extracted from the encrypted images at the CS side. The index building will be further built by CS. It should be noted that both features and the index are in the encrypted domain, however, the Manhattan distance between features is still same as the distance in plaintext state. Since the global feature is too weak for the retrieval, Xia \textit{et al.} \cite{xia2019boew} further propose a scheme called BOEW which considers the encryption in each block and ensure that the encrypted local features are valid for retrieval. The server further aggregates the encrypted local features with the typical BOW model which greatly improves retrieval efficiency.

As the encryption in spatial domain malfunctions the compression, that is, it will greatly increase the size of encrypted images. Therefore, some schemes \cite{cheng2016encrypted,liang2019huffman} try to execute PPCBIR in JPEG-domain. These methods encrypt the image in JPEG-format and ensure the JPEG-format be kept after the encryption. The encryption methods used are almost similar (e.g., permutation) to the schemes in spatial domain. It should be noted that the above schemes are all based on probabilistic encryption, which makes the security of their schemes depends on the size of image content, and the statistic feature shows a mediocre performance for retrieval.

\subsubsection{schemes based on typical feature}

The second class tries to extract typical features (e.g., SIFT) from the encrypted images. Different from the statistic feature, plenty of linear, nonlinear, and comparison computations appear during the extraction process of typical features. To our knowledge, Hsu \textit{et al.} \cite{hsu2012image} first investigate privacy-preserving SIFT extraction in the encrypted domain based on the Homomorphic Encryption (HE)\cite{gentry2009fully} tool which supports direct computation on the ciphertext. However, their scheme suffers from high interaction rounds and potential security risks. The following schemes \cite{wang2013secure} notice that one single server is hard to cope with these problems. They further try to use multiple servers (e.g., two servers) to collaboratively execute the encrypted SIFT feature extraction. For example, Hu \textit{et al.} \cite{hu2016securing} combine the Somewhat HE\cite{paillier1999public} and parallel technology, propose batched secure multiplication and comparison protocol between two servers. However, these schemes are still too time-consuming rather than practice. In recent work, Liu \textit{et al.} \cite{liu2019intelligent} try to extract encrypted features based on a pre-trained CNN model (e.g., VGG16). The better feature extractor brings much higher accuracy.

It should be noted that the encrypted feature extraction by pre-trained CNN is equivalent to the following problem: how to execute the inference process of CNN with known parameters in safety. There are many schemes \cite{huang2019lightweight} try to handle this fundamental problem in recent years. The HE tool is also considered first. However, due to the nonlinear computation or comparison in the activation layer and pooling layer, the basic HE can not satisfy demands. There are two main strategies to cope with this problem: one type of schemes \cite{juvekar2018gazelle} tries to use SMC technology, such as garbled circuit\cite{yao1982protocols}, to assist nonlinear operations; the others \cite{gilad2016cryptonets} attempt to seek approximate algorithms to replace them. Actually \cite{liu2019intelligent} can be seen as the utilization of \cite{juvekar2018gazelle}. However, these schemes not only cause plenty of time consumption but also need to convert ciphertext to an integer (or bits), which will cause extra precision loss.

To alleviate the problem, Huang \textit{et al.} \cite{huang2019lightweight} proposed a scheme completely based on additive secret sharing which naturally supports linear computation on the ciphertext. They further construct a secure comparison protocol based on $Beaver's$ $triples$\cite{beaver1991efficient}. The avoidance of HE makes the scheme obtain a great advantage in time consumption, however, the protocols they proposed still have lots of room for improvement.

In this paper, we propose a novel PPCBIR scheme based on pre-trained CNN. Different from the previous low-efficiency schemes, we first propose enough novel computation protocols based on additive secret sharing, then simulate the feature extraction in state-of-the-art CBIR schemes, and further propose secure feature compression and index building technologies to decrease the consumption. The utilization of these strategies makes our scheme both efficient and accurate.

\section{Preliminaries}\label{sec:Prelimin}

\subsection{Image feature extracted by pre-trained CNN}\label{subsec:FeatureExtractByCNN}

In this work, the feature is extracted from a typical CNN model called VGG16, and the parameters pre-trained by a large image dataset \cite{russakovsky2015imagenet} are utilized. CNN is a sequence of layers that compute and transform the input into output. Each layer is made up of a set of neurons. Like most of the other CNN models, VGG16 is composed of Fully Connected layer (FC), Convolutional layer (Conv), Activation Layer (AL), and Pooling layer (Pool). In detail, each layer contains the following computation:

\begin{itemize}[]

	\item \textbf{FC and Conv}. FC and Conv layer only involve the linear computation. For example, the neurons in Conv layer share the same weights $W$ and biases $b$, which are often called filters. Consider a filter sizes $n\times n$, then each neuron in the current layer is transformed from an $n\times n$ region of neurons in previous layer. In detail, the  $(j,k)^{th}$ neuron is obtained from $y_{j,k}$ $=$ $\sum_{l=0}^{n-1}\sum_{m=0}^{n-1}w_{l,m}x_{j+1,k+m}+b$. Here, $w_{l,m}$ means the corresponding value in weights matrix $W$ and $x$ represents the neuron in the previous layer. It should be noted that the $W$ is fixed when we use pre-trained CNN. As we disuse the FC layer in the VGG16, our scheme can deal with images in any size.

	\item \textbf{AL}. Activation layer provides the non-linear computation in the neural networks, and Rectified Linear Unit (ReLU) is the unique AL in VGG16. For the input matrix $X$, the ReLU layer returns the result $ReLU(X)$ composed by $max(x,0)$, where $x$ is the member value in $X$. It is easy to notice that ReLU only involves comparison computations.
	
	\item \textbf{Pool}. The pooling layer partitions the input layer into a set of non-overlapping rectangles, then a down-sampling operation is executed on each rectangle to get the value in the current layer. In VGG16, max-pooling sizes 2$\times$2 is utilized, which means the maximum number will be saved in each 2$\times$2 block and transmitted to the later layer. Max-pooling also only contains the comparison computations.

\end{itemize}

\subsection{Additive Secret Sharing}

Secret sharing is one of the most important technologies in SMC. The secret information will be randomly split into multi-shares, and each participant only has one share. The secret can be reconstructed only when a sufficient number of shares are combined. Additive secret sharing can be seen as an $(n,n)$-threshold secret sharing technology. It means that the secret $x$ can be randomly split into $n$ shares, for example, $x$ $=$ $x_1+x_2+\cdots+x_n$, and recover the secret needs all $n$ share, here $n\geq 2$. In this paper, we only focus on the situation that $n = 2$.

In recent years, more researchers try to use the additive secret sharing technologies to outsource computation tasks to CS. For example, the image owner can outsource his image by randomly splitting the image into two additive shares, and each cloud server $S_i$ $(i=1,2)$ owns one share. For simplicity, the following $S_i$ are all represents servers $S_1$ and $S_2$, similarly, the following values, with $i$ in subscript, represent the additive share of corresponding value if $i$ is undefined. It should be noted that the additive secret sharing has an important property: each server $S_i$ can execute linear computation without interaction. For example, each server owns $u_i, v_i$, and they want to compute $u \pm v$. Due to $(u_1$ $\pm$ $v_1)$ $+$ $(u_2$ $\pm$ $v_2)$ $=$ $u$ $\pm$ $v$, $S_i$ only needs to compute $u_i$ $\pm$ $v_i$, and the result is still the share of $u$ $\pm$ $v$. Similarly, each server can execute the secure multiplication by a known number (e.g., pre-trained parameters).

$Beaver's$ $triples$ is widely used in additive secret sharing to execute the secure multiplication called \texttt{SecMul}. During the \emph{offline phase} (i.e., before computing secure multiplication), a pre-computed triple of the form $\{a_i,b_i,c_i|ab=c\}$ is generated and $(a_i, b_i, c_i)$ is sent to server $S_i$, here $(a_i, b_i, c_i)$ is the additive share of $(a, b, c)$. In \emph{online phase}, each server computes $e_i = x_i - a_i$ and $f_i = y_i - b_i$. Then two server exchange the information of $e_i$ and $f_i$. Finally, $S_i$ can compute $z_i = fa_i + eb_i + c_i + (i-1)ef$. It is easy to notice that $z_1 + z_2 = xy$. The existing schemes \cite{wagh2018securenn} further expand the secure scalar multiplication to secure matrix dot product as shown in algorithm \ref{alg: SecMatMul}.
\begin{algorithm}[t]
	\caption{Secure matrix multiplication $\texttt{SecMatMul}$}
	\label{alg: SecMatMul}
	\begin{algorithmic}[1]
		\REQUIRE
		$S_i$ has $X_i, Y_i$.
		\ENSURE
		$S_i$ gets $(X\cdot Y)_i$.
		
		\textbf{Offline Phase} :
		\STATE $\mathcal{T}$ generates random matrix $A$, $B$ and computes $C$=$A$$\cdot$$B$.
		\STATE $\mathcal{T}$ randomly splits ($A$, $B$, $C$) into two additive share ($A_i$, $B_i$, $C_i$) and sends the share to corresponding server $S_i$.
		
		\textbf{Online Phase} :
		\STATE $S_i$ computes $E_i = X_i - A_i$ and $F_i = Y_i - B_i$.
		
		\STATE $S_1$ and $S_2$ exchange the share $E_i$ and $F_i$.
		
		\STATE $S_i$ computes $X_i$ $\cdot$ $F$ $+$ $E$ $\cdot$ $Y_i$ $+$ $C_i$ $-$ $(i-1) E\cdot F$.
		
	\end{algorithmic}
\end{algorithm}

Although the fixed size vectors need to be computed during \emph{Offline phase} in algorithm \ref{alg: SecMatMul}, we would show that it is feasible when facing a specific task in subsection \ref{subsec:Discussion on TTP}. In recent work, Xiong \textit{et al.} \cite{xiong2020efficient} further expands $Beaver's$ $triples$, and constructs protocols which can support almost all basic elementary function and four fundamental operations with only constant-rounds interaction rounds. In the rest of the paper, an additive share is called a share for simplicity.

\subsection{HKM and C2LSH}\label{subsec:HKMAndC2LSH}

Excessive comparison operations will lead to low retrieval efficiency, in this case, a reasonable index is indispensable. In PPCBIR schemes based on typical features, these methods can not only accelerate the retrieval, but also significantly reduce the communication consumption that will be shown in subsection \ref{subsec:retrievalconsum}. For simplicity, the HKM and C2LSH are chosen to represent tree-based and hash-based index building schemes. As we focus on the way which employs these technologies when feature vectors are stored as a share in two servers, only the numerical calculation process on feature vectors is given below. The readers can refer to \cite{muja2014scalable} and \cite{gan2012locality} for more details (e.g., virtual rehash) about these schemes.

\subsubsection{Index building and query of HKM}

As described in subsection \ref{subsec:CBIR}, there are mainly two steps in the index building process of HKM:

\emph{(i) Index vectors by $k$-means}. The server divides all the vectors into $k$ different categories by $k$-means first. There are three main steps during executing $k$-means: firstly, $k$ random vectors are chosen as the centroid vectors; secondly, each vector computes their distance to centroid vectors and joins in the nearest category; finally, the centroid vectors will be updated as the mean value of new vectors in the current category. The second and third steps will be executed iteratively until a fixed number of times or the centroid vectors becomes stable.

\emph{(ii) Repeat $k$-means partition}. After initial partitioning, the vectors in each subspace may still be too many. In this case, to ensure the number of vectors in each subspace less than a fixed number, the server will execute $k$-means on the corresponding subspace where excessive vectors exist. The partitioning will not end until each subspace meets the requirements.

\begin{figure}[tb]
	\centering
	\includegraphics[width=1.0\linewidth]{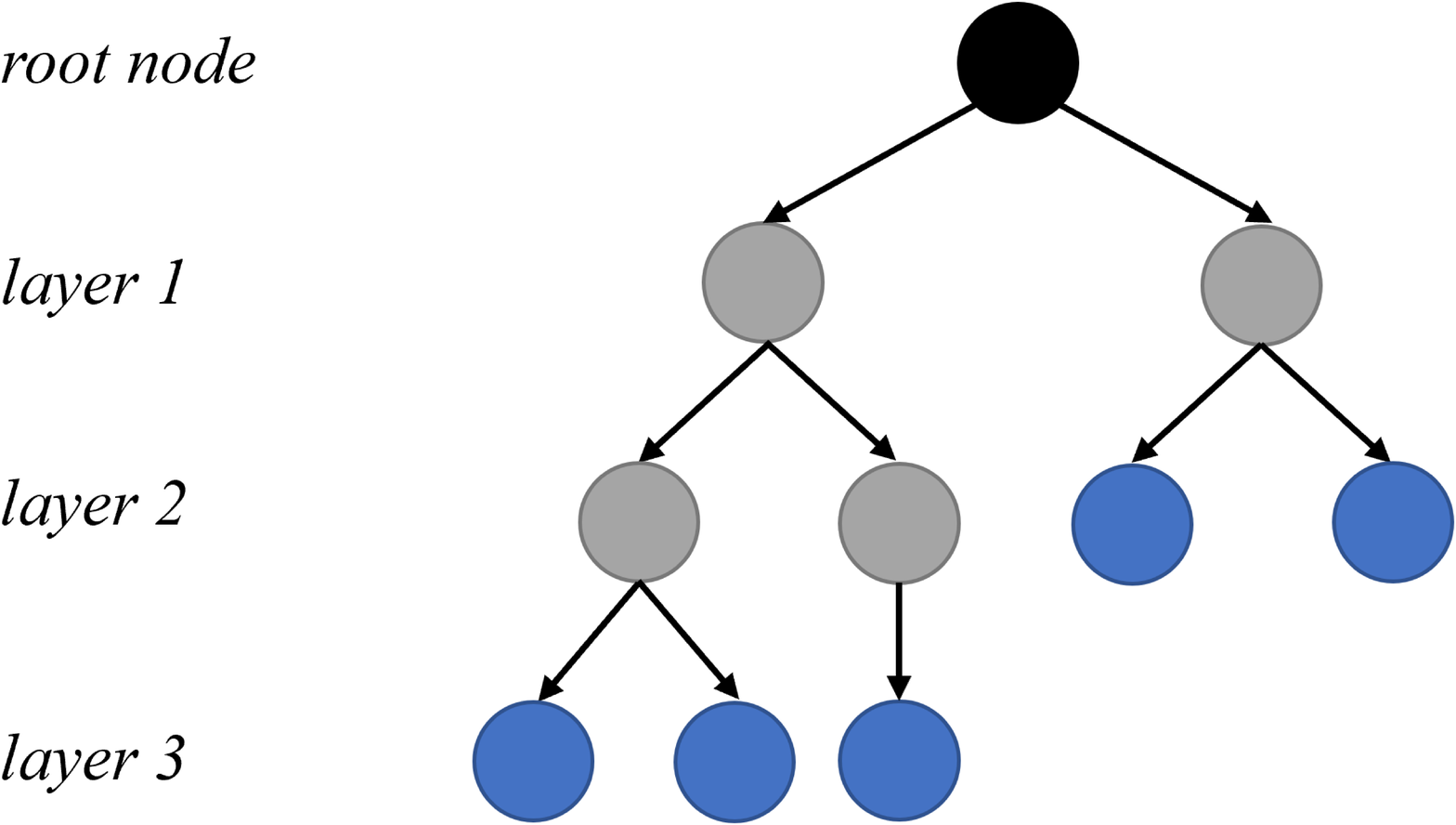}
	\caption{An example of HKM. In the figure, five feature vectors (blue nodes) are indexed by HKM where $k$ set as 2.} 
	\label{fig:HKMTree}
\end{figure}

After the index building, each vector will become one leaf node of the HKM tree like Fig. \ref{fig:HKMTree}. Here, non-leaf nodes (except the root node) are the centroids of corresponding space.

During the retrieval, to achieve high efficiency and accuracy, the server will choose enough suitable vectors to compose $candidates$. The true distances between query and $candidates$ will be computed and $m$-nearest vectors will be selected from them. Generally speaking, the number of vectors in $candidates$ is usually over $m$ (e.g., 3 times of $m$), in this case, the core challenge is how to seek enough and high-quality $candidates$. There are mainly two steps in HKM after the server gets the query vector:

\emph{(i) Seek the nearest vectors}. The server computes distances between query and centroids in the first layer. Then the server enters the subspace nearest node represents and adds all other brother nodes into the candidate list. The above process will be repeated until the server finds leaf nodes.

\emph{(ii) Get enough candidate vectors}. In HKM, the feature vectors that the leaf nodes in the above step represent will be all added into $candidates$. If not enough, the server will set the nearest node in the candidate list as the root node and repeat the above step to get new leaf nodes.

\subsubsection{Index building and query of C2LSH}

As described in subsection \ref{subsec:CBIR}, there are mainly two steps in the index building process of C2LSH:

\emph{(i) Generate hash functions}. The LSH family in C2LSH is shown as follows:

\begin{equation}
\label{eq:C2LSH}
\begin{array}{c}
h_{\vec{a},b}(o) = \lfloor \frac{h'_{\vec{a},b}(o)}{w} \rfloor,
\end{array}
\end{equation}
\begin{equation}
\label{eq:C2LSH2}
\begin{array}{c}
h'_{\vec{a},b}(o) = \vec{a}\cdot \vec{o} + bwx,
\end{array}
\end{equation}

\noindent where $\vec{o}$ is a feature vector to be mapped, $\vec{a}$ is the same dimension vector where each entry is drawn independently from the standard normal distribution $N(0,1)$, $w$ is a user-specified constant (e.g., 1), $b$ is a real number uniformly random drawn from $[0, w)$ and $x$ is a positive random integer. It is easy to notice that the hash function in C2LSH is composed of $(\vec{a},$ $b,$ $w,$ $x)$. The server will generate a series of LSH functions in this family.

\emph{(ii) Map vectors into LSH tables}. After generating the LSH functions, the server will map all the vectors in the database to a series of LSH tables. Each LSH function will generate a corresponding LSH table as shown in Table \ref{tab:C2LSHTable}. If the $\{h_{\vec{a}_j,b_j}(o)\}$ of feature vector $\vec{o}$ equal to $bid$, then the corresponding image identity will be added into bucket $bid$ in the $j$-th LSH table, here $j\in [1, m]$, and $m$ is the number of generated LSH functions.

\begin{table}[ht]
	\centering
	\caption{An example of LSH Table. Here $ID$ is the identity of the corresponding image.}
	\label{tab:C2LSHTable}
	\begin{tabular}{|c|l|}
		\hline
		Bucket identity & Image identities in current bucket \\ \hline
		$bid_1$ & $ID_{1}$, $ID_{35}$, $ID_{143}$, $ID_{146}$\\ \hline
		$bid_2$ & $ID_{3}$, $ID_{131}$, $ID_{142}$\\ \hline
		$\cdots$ & $\cdots$                      \\ \hline
		$bid_n$ & $ID_{12}$, $ID_{23}$, $ID_{138}$, $ID_{214}$, $ID_{235}$\\ \hline
	\end{tabular}
\end{table}

There are mainly two steps during retrieval based on C2LSH:

\emph{(i) Map query into LSH tables}. Same as the vectors in the database, server will use all LSH functions to map the query vector $\vec{q}$ into corresponding LSH tables. Then the vectors in the same $bid$ in any LSH table will be seen as one collusion with the query. The collision information will be stored, and the vector will be selected into $candidates$ if it reaches a certain number of collisions.

\emph{(ii) Get enough candidate vectors}. Similarly, the vectors in current $bid$ may be insufficient. C2LSH further propose a scheme called virtual rehash to seek new $bids$ for further collision count and the calculation during virtual rehash only needs $h'_{\vec{a},b}(q)$ in each LSH table. \cite{gan2012locality} gives two ways to stop seeking $candidates$: the first one stops when $m$ or more vectors whose distance from the query vector is less than a specific value are found; the second one does when enough (e.g., $3\times m$) vectors reach a certain number (related to the number of vectors in the database) of collisions, where $m$ is the number of vectors returned.

\section{Proposed System architecture}\label{sec:Sysmodel}

\subsection{System model}\label{subsec:systemmodel}

\begin{figure}[tb]
	\centering
	\includegraphics[width=1.0\linewidth]{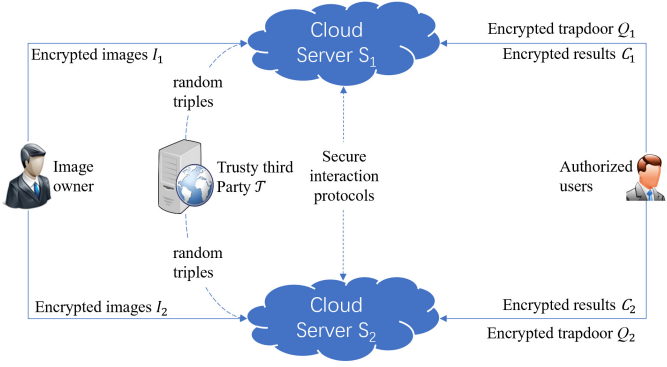}
	\caption{System model.}
	\label{fig:systemmodel}
\end{figure}

Similar to \cite{hu2016securing} and \cite{liu2019intelligent}, the proposed scheme involves five entities, i.e., the image owner, the cloud server $S_1$, cloud server $S_2$, a trusty party $\mathcal{T}$ and authorized users. The system model is shown in Fig. \ref{fig:systemmodel}.

\textbf{\boldmath{Image owner}} owns a large-scale database $\mathcal{I} = \{I_i\}_{i=1}^{n}$ with the corresponding identity set $\mathcal{ID} = \{ID_i\}_{i=1}^{n}$ to be outsourced, here $n$ represents the number of images. To preserve privacy, the images are all randomly split into two parts $\mathcal{C}_1 = \{C_i^1\}_{i=1}^{n}$ and $\mathcal{C}_2 = \{C_i^2\}_{i=1}^{n}$ based on additive secret sharing. The encrypted image databases are separately sent to $\mathcal{S}_1$ and $\mathcal{S}_2$.

\textbf{\boldmath{Cloud server}} $S_1$ and $S_2$ undertake the task of feature extraction, feature compression, index building, and secure retrieval. Two cloud servers need to interact with each other and collaboratively complete the above tasks.

\textbf{\boldmath{Trusty third party}} called $\mathcal{T}$ takes the charge of generating random numbers or random vectors which will be used during the secure computation. It should be noted that the generation of numbers is \emph{offline} and lightweight work. The $\mathcal{T}$ can be undertaken by the government or the owner of images, we will further discuss this problem in subsection \ref{subsec:Discussion on TTP}.

\textbf{\boldmath{Authorized user}} authorized by the image owner has the authorization to retrieve the corresponding images. During the retrieval, user sends the trapdoor to two servers and gets the encrypted images from the servers. The plaintext results can be obtained by simply adding the corresponding encrypted versions.

\subsection{Security Model}\label{subsec:securitymodel}

Similar to many previous works \cite{hu2016securing} in PPCBIR based on typical features, the semi-honest but non-collusion CS is considered in our scheme. It means each server will execute the protocol as asked but will try to analyze the information from data, however, due to their reputation and financial interests, they will not collude with each other. As described above, the third party $\mathcal{T}$ only needs to generate random numbers in the \emph{offline phase}, which means it can be undertaken by a low-computation device. Therefore, it is reasonable to assume that $\mathcal{T}$ is honest and non-colluding.

\section{Basic secure computation based on two server}\label{sec:secretMatrixComputation}

\subsection{Secure Comparison}\label{subsec:secureCmp}

The typical comparison schemes \cite{damgaard2006unconditionally,huang2019lightweight} based on additive secret sharing try to use the most significant bit; however, it will lead to $l$ rounds interaction, where $l$ is the bit-length of ciphertext. These schemes try to ensure the information-theoretic security of input; however, in the computation process of a practical task, plenty of intermediate values are meaningless. It means the leakage of some non-sensitive  information (e.g., the range of numbers) will cause no actual damage to the specific task. In this paper, to avoid the difficulty of comparison, we try to get the sign of $secret$ by transforming the number into an irreversable number with the same sign.

Inspired by \cite{liu2019toward}, the comparison protocol based on two servers can be constructed shown as algorithm \ref{alg: SecureComparison}. Each server firstly computes the share of $u-v$, then a random number $r$ generated by $\mathcal{T}$ is utilized to cover the $u-v$. In this paper, the $sgn(x)$ is defined as formula \ref{eq:sgn}. If $r$ is positive, then the $share$ of $1$ will be sent to servers; otherwise, the $share$ of $0$ will be sent. Notably, since the multiplication on $0$ will result $0$ naturally, the shares of sign can help us execute ReLU or max-pooling functions secretly with the help of $\texttt{SecMul}$.

\begin{equation}
\label{eq:sgn}
\begin{array}{c}
sgn(x) = 
\left\{
\begin{aligned}
1\qquad & x > 0	\\
0\qquad & otherwise
\end{aligned}
\right.

\end{array}
\end{equation}

To enhance the randomness, a small random number $k$ is added. If the result $f=f_1+f_2$ is positive, it implies that $u-v$ is also positive, and vice versa. Please note that, although $k$ may cause the wrong judgment; however, on the one hand, it can hide the situation that $u$ equals to $v$. On the other hand, for an actual task, the result of the comparison is generally used for deciding a choice. The adjacent values always imply both the choices are feasible. For instance, if two images have similar distances to the query, it strongly implies that the two images are both the wanted or not wanted results.

\begin{algorithm}[t]
	\caption{Secure comparison protocol \texttt{SecCmp}}
	\label{alg: SecureComparison}
	\begin{algorithmic}[1]
		\REQUIRE
		$S_i$ has $u_i$, $v_i$.
		\ENSURE
		$S_i$ gets $sgn(u-v)_i$.
		
		\textbf{Offline Phase} :
		
		\STATE $\mathcal{T}$ generates a random nonzero number $r$, then computes the share $r_i$ and the share of the $sgn(r)$, then sends them to $S_i$.
		
		\STATE $\mathcal{T}$ generates a random number $k$, where $|k|$ $\ll$ $|r|$, then computes the share $k_i$ and sends them to $S_i$.
		
		\STATE $\mathcal{T}$ generates enough random numbers that the sub-protocol uses and sends them to $S_i$.
		
		\textbf{Online Phase} :
		
		\STATE $S_i$ computes $t_i = u_i - v_i$.
		
		\STATE $S_1$ $\&$ $S_2$ collaboratively compute $f_i$ = $\texttt{SecMul}(t_i, r_i) + k_i$.
		
		\STATE $S_1$ $\&$ $S_2$ collaboratively recover $f$. If $f$ is positive, $sgn(u-v)_i = sgn(r)_i$; else, $sgn(u-v)_i = \frac{1}{2} - sgn(r)_i$.
		
	\end{algorithmic}
\end{algorithm}

We will analyze the security and potential information leakage of this protocol in section \ref{sec:SecuAnal}. With the help of $share$ of sign, it is easy to utilize $\texttt{SecCmp}$ to compute the ReLU function as shown in algorithm \ref{alg: SecureRELU}. Compared to previous works \cite{huang2019lightweight}, the sign of elements in ReLU layer is also under the protection in our work.

\begin{algorithm}[t]
	\caption{Secure ReLU function}
	\label{alg: SecureRELU}
	\begin{algorithmic}[1]
		\REQUIRE
		$S_i$ has $x_i$.
		\ENSURE
		$S_i$ gets $ReLU(x)_i$.
		
		\textbf{Offline Phase} :
		
		\STATE $\mathcal{T}$ generates enough random numbers that the sub-protocols use and sends them to $S_i$.
		
		\textbf{Online Phase} :
		
		\STATE $S_1$ $\&$ $S_2$ collaboratively compute $sgn(x)_i$ $=$ $\texttt{SecCmp}($ $x_i, 0)$.
		
		\STATE $S_1$ $\&$ $S_2$ collaboratively compute $ReLU(x)_i$ $=$ $\texttt{SecMul}($ $x_i,$ $sgn(x)_i)$.
		
	\end{algorithmic}
\end{algorithm}

\subsection{Secure Matrix Inverse}

The motivation of $\texttt{SecCmp}$ is transforming the number, which needs protection, to another random number in safety and then re-share the new number. At the same time, the complex computation we execute on the new number can be reflected in the original number after a reasonable transform. In this way, the complexity of operations like comparison can be avoided. This motivation can be further applied to the matrix computation. Inspired by \cite{bar1989non}, we propose the secure matrix inverse protocol as algorithm \ref{alg: SecMatInv}.

\begin{algorithm}[t]
	\caption{Secure matrix inversion protocol $\texttt{SecMatInv}$}
	\label{alg: SecMatInv}
	\begin{algorithmic}[1]
		\REQUIRE
		$S_i$ has $X_i$.
		\ENSURE
		$S_i$ gets $Y_i = (X^{-1})_i$.
		
		\textbf{Offline Phase} :
		
		\STATE $\mathcal{T}$ generates enough random matrices the sub-protocol uses and sends to $\mathcal{S}_i$.
		
		\textbf{Online Phase} :
		
		\STATE $S_i$ generates a random square matrix $Z_i$.
		
		\STATE $S_1$ $\&$ $S_2$ collaboratively compute $W_i$ $=$ $\texttt{SecMatMul}($$Z_i,$ $X_i)$.
		
		\STATE $S_1$ $\&$ $S_2$ re-share and get $W$, then $S_i$ computes $W^{-1}$.
		
		\STATE $S_i$ computes $Y_i$ = $W^{-1} \cdot Z_i$.
		
	\end{algorithmic}
\end{algorithm}

In \texttt{SecMatInv}, each server first generates a random square matrix, then servers compute its dot product with the input. As the property that $(Z\cdot X)^{-1}\cdot (Z_1+ Z_2) = X^{-1}$, each server can finally get the share of $X^{-1}$. As the sum of the random matrix generated by two servers is almost certainly (i.e., the probability is 1) reversible, for simplicity, we here ignore the potential risk brought by the random matrix. It should be noted that this situation can be detected by calculating the rank of $W$. Especially when the matrix is simplified as a number, the secure division computation $\texttt{SecDiv}$ is shown as algorithm \ref{alg: SecDiv}.

\begin{algorithm}[t]
	\caption{Secure division protocol $\texttt{SecDiv}$}
	\label{alg: SecDiv}
	\begin{algorithmic}[1]
		\REQUIRE
		$S_i$ has $u_i$ and $v_i$.
		\ENSURE
		$S_i$ gets $(\frac{u}{v})_i$.
		
		\textbf{Offline Phase} :
		
		\STATE $\mathcal{T}$ generates enough random numbers the sub-protocol uses and sends to $S_i$.
		
		\textbf{Online Phase} :
		
		\STATE $S_i$ generates random number $r_i$.
		
		\STATE $S_1$ $\&$ $S_2$ collaboratively compute $f_i$ = $\texttt{SecMul}(u_i, r_i)$ and $g_i$ = $\texttt{SecMul}(v_i, r_i)$.
		
		\STATE $S_1$ $\&$ $S_2$ re-share and get $g$, then $S_i$ computes $\frac{u_i}{g}$.
		
	\end{algorithmic}
\end{algorithm}

\section{Privacy-preserving CBIR based on two servers}\label{sec:featurextrac}
\begin{figure*}[tb]
	\centering
	\includegraphics[width=1.0\linewidth]{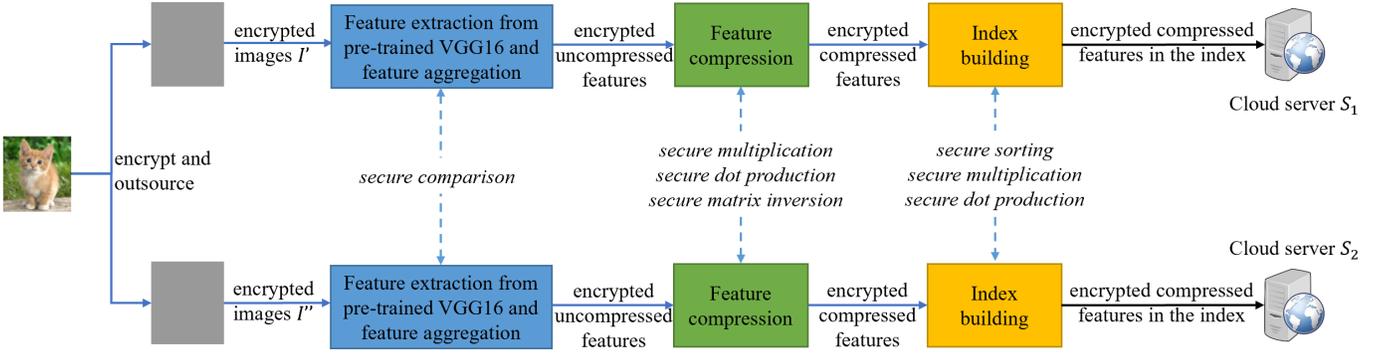}
	\caption{An overview of image outsourcing. The cloud servers will collaboratively compute and store the encrypted compressed features in the index}
	\label{fig:entireProcess}
\end{figure*}

In this section, we firstly give the feature extraction process based on additive secret sharing, then further show the PCA compression method on encrypted features. Finally, the secure index building methods based on HKM and C2LSH are given for better retrieval efficiency. An overview of the process is shown in Fig. \ref{fig:entireProcess}. 

\subsection{Secure Pre-trained CNN Feature extraction and aggregation}\label{subsec:FeaExtra}

In this paper, following the suggestion in \cite{uijlings2013selective}, the last activation layer before \emph{FC} in pre-trained VGG16 is utilized as a feature extractor. As subsection \ref{subsec:systemmodel} shows, the CNN model this paper involves mainly contains three different types of layers called \emph{Conv}, \emph{AL}, and \emph{Pool}.

As the parameters or hyper-parameters are fixed and known by both two servers, which means all the weights are just constant in the CNN. For \emph{Conv}, it is just the secure constant multiplication which means no interactions are necessary during the inference process.

ReLU is the unique $\emph{AL}$ in VGG16. As described in subsection \ref{subsec:secureCmp}, the number less than $0$ will be set as $0$ secretly with the help of the share of $0$. Benefiting from parallelism, only three rounds of interaction are needed between the servers.

The max-pooling layer in VGG16 needs to seek the position of the maximum value in each $2\times 2$ block. Similar to ReLU function, the protocol for max-pooling is shown in algorithm \ref{alg: MaxPool}. The smaller value will be set as 0 secretly and only the greater one will be reshared. Benefiting from parallelism, only 6 rounds of interaction are needed in the max-pooling layer.

\begin{algorithm}[t]
	\caption{Secure max-pooling in $2\times{}2$ block}
	\label{alg: MaxPool}
	\begin{algorithmic}[1]
		\REQUIRE
		$S_i$ has $x^1_i$, $x^2_i$, $x^3_i$, $x^4_i$.
		\ENSURE
		$S_i$ gets $max(x^1,x^2,x^3,x^4)_i$.
		
		\textbf{Offline Phase} :
		
		\STATE $\mathcal{T}$ generates enough random numbers that the sub-protocols use and sends them to $S_i$.
		
		\textbf{Online Phase} :
		
		\STATE $S_1$ $\&$ $S_2$ collaboratively compute $sgn(x^{12})_i$ $=$ $\texttt{SecCmp}(x^1_i, x^2_i)$; $sgn(x^{34})_i$ $=$ $\texttt{SecCmp}(x^3_i, x^4_i)$.
		
		\STATE $S_1$ $\&$ $S_2$ collaboratively compute $max(x^1,x^2)_i = \texttt{SecMul}(sgn(x^{12})_i, x^1_i) + \texttt{SecMul}((\frac{1}{2}-sgn(x^{12})_i), x^2_i)$; $max(x^3,x^4)_i = \texttt{SecMul}(sgn(x^{34})_i, x^3_i) + \texttt{SecMul}((\frac{1}{2}-sgn(x^{34})_i), x^4_i)$.
		
		\STATE $S_1$ $\&$ $S_2$ collaboratively compute $sgn(x^{1234})_i$ $=$ $\texttt{SecCmp}(max(x^1,x^2)_i, max(x^3,x^4)_i)$.
		
		\STATE $S_1$ $\&$ $S_2$ collaboratively compute $max(x^1,x^2,x^3,x^4)_i = \texttt{SecMul}(sgn(x^{1234})_i, max(x^1,x^2)_i) + \texttt{SecMul}((\frac{1}{2}-sgn(x^{1234})_i), max(x^3,x^4)_i)$.
		
	\end{algorithmic}
\end{algorithm}

After the extraction, following the conclusion in \cite{babenko2015aggregating}, the average aggregation is used to aggregate all the numbers gotten by each filter. Obviously, no interaction is needed during the aggregation. Here, each server gets a 512-dim encrypted feature vector for each image.

\subsection{Secure Feature compression}\label{subsec:FeaCompress}

PCA is the most commonly used feature compression method\cite{wei2017selective} for accelerating the retrieval. Here, we try to propose secure PCA computation on the encrypted features collaboratively stored in two servers. Note that the compression can significantly reduce communication consumption during secure retrieval.

To our knowledge, no existing schemes try to execute the secure PCA based on additive secret sharing, the essential problem here is the way in computing eigenvectors of the matrix in safety. The following Lemmas should be noticed in this case:

\textbf{Lemma 1.} \emph{If $A$ and $B$ are similar matrices (i.e., $A\sim B$), and $B=P^{-1}AP$, then $A$ and $B$ have the same eigenvalues; If there is an eigenvector $\vec{x}$ under eigenvalue $\lambda$ of matrix $A$, then $P^{- 1}\vec{x}$ is an eigenvector of $B$ under eigenvalue $\lambda$.}

\textbf{Lemma 2.} \emph{If $\lambda$ is an eigenvalue of matrix $A$, then $k\times \lambda$ is an eigenvalue of matrix $k\times A$; If there is an eigenvector $\vec{x}$ under eigenvalue $\lambda$ of matrix $A$, then the eigenvector $\vec{x}$ is also under eigenvalue $k\times \lambda$ of matrix $k\times A$.$(k\neq{}0)$}

\begin{algorithm}[t]
	\caption{Secure PCA protocol}
	\label{alg: SecPCA}
	\begin{algorithmic}[1]
		\REQUIRE
		$S_i$ has $X_i^{n\times d}$ $(n>>d)$.
		\ENSURE
		$S_i$ gets $F_i^{n\times s}$.
		
		\textbf{Offline Phase} :
		
		\STATE $\mathcal{T}$ generates enough random numbers and matrices the sub-protocols use and sends to $\mathcal{S}_i$.
		
		\textbf{Online Phase} :
		
		\STATE $S_i$ computes the mean value for each column of $X_i$ and composes $d$-dim vector $\vec{m}_i$, then computes $X_i$ = $X_i$ - $\vec{m}_i$.
		
		\STATE $S_i$ generates a random square matrix $P^{d\times d}_i$ and a random number $t_i$.
		
		\STATE $S_1$ $\&$ $S_2$ collaboratively compute $P^{-1}_i$ = $\texttt{SecMatInv}(P_i)$.
		
		\STATE $S_1$ $\&$ $S_2$ collaboratively compute $Y_i$ $=$ $\texttt{SecMul}(t_i,$ $\texttt{SecMatMul}(P^{-1}_i,$ $\texttt{SecMatMul}($$\texttt{SecMatMul}($$X_i^T,$ $X_i),$ $P_i)))$.
		
		\STATE $S_2$ sends $Y_2$ to $S_1$, then $S_1$ computes the eigenvectors of $Y$, then pick the $T^{d\times s}$ composed by eigenvectors under the $s$ maximal eigenvalues and sends them to $S_2$.
		
		\STATE $S_i$ computes $V_i^{d\times s}$ = $P_i \cdot T$.
		
		\STATE $S_1$ $\&$ $S_2$ collaboratively compute $U_i^{d\times s}$ which is composed by $\texttt{SecMul}(v^{jk}_i,$ $v^{jk}_i)$ for all $v^{jk}_i$ in $V_i$, here $j\in [1,d],$ $k\in [1,s]$.
		
		\STATE $S_i$ computes the sum value for each column in $U_i^{d\times s}$, and compose a $s$-dim vector $\vec{r}_i$, then re-share it.

		\STATE $S_i$ gets $\vec{r}$ and computes $v^{jk}_i = \frac{v^{jk}_i}{\sqrt{r_k}}$ for all $v_{i}^{jk}$ in $V_i$, here $r_k$ means the $k$-th element in $\vec{r}$.
		
		\STATE $S_1$ $\&$ $S_2$ collaboratively compute $F_i^{n\times s}$ = $\texttt{SecMatMul}(X_i, V_i)$.
		
	\end{algorithmic}
\end{algorithm}

Based on Lemma 1 and 2, algorithm \ref{alg: SecPCA} shows the scheme of PCA on an encrypted matrix. Each server first executes zero-centering for each column of the matrix $X_i^{n\times d}$, the $\vec{m}_i$ composed by mean values will be stored in the server.

Then, each server will randomly generate a matrix $P_i$ and a random positive number $t_i$, and the matrix $(P^{-1})_i$ will be collaboratively computed based on $\texttt{SecMatInv}$. Here the $P$ and $P^{-1}$ are generated to transform the original matrix, and $t$ is used to protect the equivalent eigenvalues. The $Y$ can be computed with the above information as shown in line number $5$. To simplify the calculation, here we only let $S_1$ undertake the task of computing eigenvectors. Due to the high concurrency during the actual task, it is easy to balance the workload on two servers.

Since $Y$ shares the same eigenvalues with $t\times X$ and $t$ is a positive number, the eigenvectors $T$ under the $s$ maximal eigenvalues of $Y$ are related to those of $X$. In this case, based on Lemma 1, the $P\cdot T$ is the corresponding eigenvectors of $X$. In line number $8$ to $10$, to ensure the stability of the results, the vectors in all $s$-dim are standardized. To reduce the interaction, the $\vec{r}$ composed by the squared sum of each eigenvector which is insensitive will be directly re-shared by servers.

Due to $\texttt{SecMul}(t_i,$ $\texttt{SecMatMul}($$\texttt{SecMatMul}($$X_i^T,$ $X_i),$ $P_i))$ can be calculated in parallel with $\texttt{SecMatInv}(P_i)$. The interaction rounds needed in secure PCA protocol is 8 if $S_1$ undertakes the task of computing eigenvectors. The complexity of communication sizes is $O(nd)$, where $n$ and $d$ are dimensions of the uncompressed matrix.

Finally, the server can get the share of compressed data based on the share of original data and compression matrix $V$. Each server will save the mean vector $\vec{m}_i$ and the compression matrix $V_i$ for the retrieval process.

\subsection{Secure Index Building}\label{subsec:IndexBuild}

To accelerate the retrieval process, the index building schemes are considered on encrypted feature vectors stored in two servers. The typical HKM and C2LSH are used as examples, and it will be easy to notice that the other schemes can be combined with our scheme too.

The comparison of distance is indispensable during the index building or query process. However, if we use the $\texttt{SecCmp}$ to compare plenty of distances and try to find the minimum one, it will lead to too high interaction rounds or too much parallel calculation. For example, to seek the minimum value in $10$ numbers, it will lead to $55$ times of comparison in parallel. Note that the size relation of distances between features is insensitive, we propose the algorithm \ref{alg: SecSort} to execute secure sorting effectively.

\begin{algorithm}[t]
	\caption{Secure sorting $\texttt{SecSort}$}
	\label{alg: SecSort}
	\begin{algorithmic}[1]
		\REQUIRE
		$S_i$ has numbers $\{u_i^l\}$, $l\in [1,n]$, $n$ is the amount of numbers.
		\ENSURE
		$S_i$ gets the sorted numbers $\{v_i^l\}$.
		
		\textbf{Offline Phase} :
		
		\STATE $\mathcal{T}$ generates a random positive number $t$, then computes the additive share $t_i$ and sends them to $S_i$.
		
		\STATE $\mathcal{T}$ generates a random number $k$, then computes the additive share $k_i$ and sends them to $S_i$ 
		
		\STATE $\mathcal{T}$ generates enough random numbers and matrices that the sub-protocols use and sends them to $S_i$.
		
		\textbf{Online Phase} :
		
		\STATE $S_1$ $\&$ $S_2$ collaboratively compute $f_i^l$ $=$ $\texttt{SecMul}(t_i,$ $u_i^l) + k_i$ for each $l$ in range $[1,n]$.
		
		\STATE $S_1$ $\&$ $S_2$ collaboratively recover the $\{f_i^l\}$, then sort $\{f^l\}$.
		
		\STATE $S_i$ gets $\{v_i^l\}$ by permuting $\{u_i^l\}$ based on the size relation of numbers in $\{f^l\}$.
		
	\end{algorithmic}
\end{algorithm}

In algorithm \ref{alg: SecSort}, each server generates a random positive number and gets $\{f_i^k\}$ by executing $\texttt{SecMul}$ on the number and share of true distances, then re-share them. In this case, the true distances will be protected by a random multiplication, but the size relation will be kept. The server can get the right size relationship of the share based on $\{f^k\}$. For efficiency, we also use the $\texttt{SecSort}$ protocol in the pooling layer for each 2$\times$2 block. After constructing the secure sorting protocol, the secure index building schemes are shown as follows.

\subsubsection{secure index building of HKM}

As described in subsection \ref{subsec:HKMAndC2LSH}, there are mainly two steps in the secure index building of HKM:

\emph{(i) Index encrypted vectors by secure $k$-means}. The key problem in this step is the way of executing secure $k$-means based on the share stored in two servers. It should be noted that the identities of images are owned by both servers, which means they can execute the calculation on the corresponding vectors independently and synchronously by a determined algorithm. At the same time, some hyper-parameters, like the number of images or the value of $k$ that HKM uses, are also known by both servers. In this case, since the operations on vectors during $k$-means are only addition and comparison, and those are easy to execute as shown in algorithm \ref{alg: SecKMeans}.

In algorithm \ref{alg: SecKMeans}, $S_1$ first randomly choose initial centroids and share it to $S_2$ for synchronization. Then, each server collaboratively executes clustering. The squared Euclidean distances are computed and $\texttt{SecSort}$ is utilized for seeking the nearest centroid. Since the vectors in each category are known by both servers, the new mean value (i.e., centroid vector) can be easily computed without interaction.

\begin{algorithm}[t]
	\caption{Secure $k$-means protocol}
	\label{alg: SecKMeans}
	\begin{algorithmic}[1]
		\REQUIRE
		$S_i$ has $k$, one share of vectors $\{\vec{x}_i^l\}$ and the $ID$ each vector corresponds, here $l\in [1,n]$, $n$ is the number of vectors.
		\ENSURE

		$S_i$ gets one share of cluster centers, and the information the vectors in each cluster.
		
		\textbf{Offline Phase} :
		
		\STATE $\mathcal{T}$ generates enough random numbers and matrices the sub-protocols use and sends to $S_i$.
		
		\textbf{Online Phase} :
		
		\STATE $S_1$ randomly picks $k$ vectors as the centroid $\{\vec{CL}_i\}$ and sends the corresponding $ID$ to $S_2$.

		\FOR{$l = 1:n$}
		
		\STATE $S_1$ $\&$ $S_2$ collaboratively compute the squared euclidean distances between $\vec{x}_i^l$ and centroid vectors based on $\texttt{SecMatMul}$.
		
		\STATE $S_1$ $\&$ $S_2$ collaboratively seek the nearest centroid vector of $\vec{x}_i^l$ based on $\texttt{SecSort}$.
		
		\STATE $S_i$ puts the $\vec{x}_i^l$ in the category which nearest centroid represents.
		
		\ENDFOR
		
		\STATE $S_i$ computes the mean value of vectors in each category, which is actually the share of new centroid vectors $\vec{CL}_i$.
		
		\STATE Repeat line 3 to line 8 until a certain rounds or $\{\vec{CL}_i\}$ unchanging.
		
	\end{algorithmic}
\end{algorithm}

\emph{(ii) Repeat secure $k$-means partition}. The situation is same as that in plaintext, as $ID$ is known by both servers, which means each server can judge and repeat partition independently and synchronously. After the index building, each server can generate the same tree structure which is also same as that in plaintext state, however, the non-leaf nodes (except root node) and leaf nodes are only the share of true values.

Since the $k$-means in each layer can be executed in parallel, the complexity of interaction rounds during HKM index building is $O(log_{k}N)$, and the complexity of communication sizes is $O(Ndlog_{k}N)$, where $k$ is the hyper-parameter in HKM, $N$ and $d$ are the number and dimension of vectors in the database.

\subsubsection{secure index building of C2LSH}\label{subsec:SecC2LSHIndexBuild}

As described in subsection \ref{subsec:HKMAndC2LSH}, there are mainly two steps during the secure index building process:

\emph{(i) Generate secure hash functions}. It should be noted that the generator of LSH function is generating random numbers that are subject to a specific distribution. In this case, the following lemma should be noted.

\textbf{Lemma 3:} \emph{If two independent random variables X, Y satisfy normal distribution $N(a_1, b_1)$ and $N(a_2, b_2)$, then X+Y satisfy normal distribution $N(a_1+a_2, b_1+b_2)$}

As C2LSH needs the $\vec{a}$ whose elements obey $N(0, 1)$, in this case, each server can generate the elements in $\vec{a}_i$ which obey $N(0, 1/2)$ independently. $w$ is a constant value known by both two servers. As $b$ is a random number in [0, $w$), therefore, it can be generated by any server. For simplicity, let us assume that this task is undertaken by $S_1$, which means $b_2 = 0$.

\emph{(ii) Mapping encrypted vectors into LSH tables}. Similar to formula \ref{eq:C2LSH2}, the servers first compute the share of $h'(o)$ collaboratively as follows,

\begin{equation}
\label{eq:SecC2LSH}
\begin{array}{c}
h'_{\vec{a},b}(o)_i = \texttt{SecMatMul}(\vec{a}_i\cdot \vec{o}_i) + b_{i}wx.

\end{array}
\end{equation}

\noindent Here $\vec{a}_i$, $\vec{o}_i$ and $b_i$ are the share of $\vec{a}$, $\vec{o}$ and $b$, the $\vec{a}$, $\vec{o}$, $b$, $w$ and $x$ are the same definition as formula \ref{eq:C2LSH2}. After computation, the servers will re-share them and get the $\{h'_{\vec{a}_j,b_j}(o^k)\}$, where $j\in [1,m]$ and $k\in [1,n]$, $m$ is the number of generated LSH functions, $n$ is the number of vectors in the database. Then each server can further compute $bid$ each share of vectors belong based on formula \ref{eq:C2LSH}. Finally, each server can add all $IDs$ into the corresponding bucket in LSH tables. After the index building, two servers will generate the same LSH tables that are also same as that in plaintext state, however, the vectors corresponding to the $IDs$ are only the share of true values.

Since the computation and share of $\{h'_{\vec{a}_j,b_j}(o^k)\}$ can be executed in parallel, the interaction rounds during C2LSH index building is $2$, and the size relation is about $O(Nd)$, here $N$ and $d$ are the number and dimension of vectors in the database.

\subsection{Secure Retrieval}\label{subsec:SecRetrieval}

This section introduces the actual retrieval process and that is the only \emph{online phase} from the respective of the secure retrieval task. The authorized user splits the query image into two shares and sends it as the trapdoor to $S_i$.

After verifying the authorization of the user, servers will extract the aggregation feature of the query as subsection \ref{subsec:FeaExtra}. Then servers collaboratively compress the extracted feature $\vec{q}_i$, with the $\vec{m}_i$ and $V_i$ stored during compression, by computing $\vec{q}_i=\texttt{SecMatMul}(\vec{q}_i-\vec{m}_i,V_i)$. Finally, according to the method of index building, servers collaboratively search the $m$ nearest images of the query. The retrieval process of HKM and C2LSH is shown as follows.

\subsubsection{secure retrieval based on HKM}\label{subsec:SecRetriHKM}

As described in subsection \ref{subsec:HKMAndC2LSH}, there are mainly two steps during the secure retrieval based on HKM:

\emph{(i) Seek the nearest encrypted vectors}. The servers will collaboratively compare the distances between query and centroids in each layer to find the nearest vectors and maintain the candidate list. The above operation only involves multiplication and sorting, which means the servers can easily finish the above task based on $\texttt{SecMatMul}$ and $\texttt{SecSort}$. The above process can be executed iteratively until servers find the leaf nodes.

\emph{(ii) Get enough encrypted candidate vectors}. The server will add the corresponding encrypted vectors in the current subspace into $candidates$. Based on the candidate list and protocols, the servers can execute the above step until they find enough $candidates$ independently and synchronously. Finally, the squared Euclidean distances will be computed based on $\texttt{SecMatMul}$, and $m$ nearest vectors can be found with the help of $\texttt{SecSort}$. The servers can finally send the corresponding encrypted images to the authorized user.

Since the servers need to seek the leaf nodes and compute the true distances between query and $candidates$, the complexity of interaction rounds during retrieval of HKM is $O(log_{k}N+3)$, the complexity of communication sizes is $O(dklog_{k}N+md)$, where $k$ is the hyper-parameter in HKM, $N$ is the number of vectors in the database, $d$ is the dimension of query vector, $m$ is the number of vectors in $candidates$.

\subsubsection{secure retrieval based on C2LSH}\label{subsec:SecRetriC2LSH}

As described in subsection \ref{subsec:HKMAndC2LSH}, there are mainly two steps during the secure retrieval based on C2LSH:

\emph{(i) Map encrypted query into LSH tables}. Similar to the calculation during the index building, the $\{h'_{\vec{a}_j,b_j}(q)_i\}$ will be computed based on encrypted query feature $\vec{q}_i$ and pre-generated encrypted LSH functions, here $j$ is the same definition as that in subsection \ref{subsec:SecC2LSHIndexBuild}. Then the $\{h'_{\vec{a}_j,b_j}(q)_i\}$ values are shared and the $bid$ of query in each LSH table can be computed. In this case, since $ID$ is known by both servers, the collision information can be counted by each server independently and synchronously.

\emph{(ii) Get enough encrypted candidate vectors}. Also, $\{h'_{\vec{a}_j,b_j}(q)_i\}$ is shared by both servers, the virtual rehash in each LSH table can be completed by each server without interaction. Like HKM, after getting enough $candidates$, the $\texttt{SecMatMul}$ and $\texttt{SecSort}$ will be finally used to get $m$-nearest images. To avoid the extra interaction, only the second completion condition in \cite{gan2012locality} is used in our experiment.

Since the servers need to map the query and compute some true distances, the interaction rounds during C2LSH index building is $5$, and the complexity of communication sizes is $O(md)$, where $d$ is the dimension of the query vector, $m$ is the number of vectors in $candidates$.

\section{Security analysis}\label{sec:SecuAnal}

Similar to previous works in PPCBIR, the security of image contents and image features are discussed in this section. We first analyze the security of the protocols constructed, then show that it is reasonable to utilize them for the PPCBIR task. For consistency, we will use the symbol defined in above sections.

\subsection{Security of protocols}

The security analysis of our scheme is under the typical secure multiparty computation framework\cite{canetti2001universally}. The execution of our scheme mainly involves the interactions between two cloud servers, and the process of the interaction is defined as the real experiment. In the proposed scheme, two cloud servers are the potential adversaries. To prove the security of the protocol, it suffices to show that the view of the corrupted party (i.e., $S_1$ or $S_2$) is simulatable given its input and output.

In the ideal experiment, the simulator $\mathcal{S}$ is defined as the one that can simulate the view of the corrupted party by using functionality $\mathcal{F}$. In this paper, similar to previous works\cite{mohassel2017secureml}, we define the $\mathcal{F}$ as follows: $\mathcal{F}$ owns the input information gotten from $\mathcal{S}_i$ and generates the random numbers or matrices locally, then it completes the calculation as subprotocols, the corresponding view of $\mathcal{S}$ will be filled by the calculation results. In the following, we will prove that the $view$ is indistinguishable from that in the real world, and the $view$ will not expose the detailed input information. To simplify the proofs, the following Lemmas will be used.

\noindent \textbf{Lemma 4. \cite{bogdanov2008sharemind}} \emph{A protocol is perfectly simulatable if all its sub-protocols are perfectly simulatable.}

\noindent \textbf{Lemma 5. \cite{xia2020secure}} \emph{The element $x+r$ is uniformly distributed and independent from $x$ for any element $x\in\mathbb{R}$ if the element $r\in\mathbb{R}$ is also uniformly distributed and independent from $x$.}

\noindent \textbf{Lemma 6. \cite{xia2020secure}} \emph{The nonzero element $xr$ is uniformly distributed and independent from $x$ for any element $x\in\mathbb{R}\backslash{}\{0\}$ if the element $r\in\mathbb{R}\backslash{}\{0\}$ is also uniformly distributed and independent from $x$.}


\textbf{Theorem 1.} \emph{The protocol $\texttt{SecCmp}$ is secure in the honest-but-curious model.}

\emph{Proof.} For $S_1$, the $view_1$ during executing protocol will be $(r_1,$ $sgn(r)_1,$ $a_1,$ $b_1,$ $c_1,$ $k_1,$ $\alpha_2,$ $\beta_2,$ $f_1,$ $f)$. Here $(r_1, sgn(r)_1, a_1,$ $b_1,$ $c_1,$ $\alpha_2,$ $\beta_2,$ $f_1)$ is random and simulatable based on Lemma 5. For simplicity, supposing $f$ is positive. Since $f$ is the result of $(u-v)$ $\times{}t$ computed by $\mathcal{F}$, based on Lemma 6, $f$ is randomly chosen from the range $(0,$ $+\infty)$. In this case, $f$ is indistinguishable from the real world. The $view_2$ for $S_2$ is similar and simulatable. $\hfill\qedsymbol$

Even if $f$ is happened to be zero, the servers can infer that $u$ and $v$ are close; however, they can not infer the detailed size relationship in that the sign of $k$ is unknown. Based on Lemma 4, the proofs for algorithm \ref{alg: SecureRELU} and \ref{alg: MaxPool} are similar and we omit them for simplicity.

\textbf{Theorem 2.} \emph{The protocol $\texttt{SecMatInv}$ is secure in the honest-but-curious model.}

\emph{Proof.} For $S_i$, except those brought by $\texttt{SecMatMul}$, the $view_i$ $=$ $(Z_i, W)$. Here $Z_i$ is composed of random numbers that are indistinguishable. $W$ is the result of $(Z_1+Z_2)$ $\cdot$ $(X_1+X_2)$ computed by $\mathcal{F}$. Due to the randomization of $Z$, any non-singular matrix in corresponding size is the potential result, which means $W$ is indistinguishable. $\hfill\qedsymbol$

When $W$ is not full-rank, similarly, any full-rank matrix in corresponding size is the potential input. Therefore, the rare situation will not bring extra risk, and the servers only need to generate $Z$ again to get the inversion.

\subsection{Security of image}

Besides the above protocols, we also construct some secure computation protocols for the PPCBIR task. For simplicity, the functionality $\mathcal{F}$ and the corresponding information leakages of these protocols are summarized in Fig. \ref{fig:FunctionLeakage}.

It is easy to note that only some insensitive intermediate values and the similarity relationship between the images are exposed. The detailed information of image contents and their features are always in safety when two servers are not colluding. To avoid almost all the information leakage, it is feasible to use \texttt{SecCmp} only; however, it will significantly decrease the efficiency of retrieval.

\begin{figure*}[ht]
	\centering
	\fbox{%
		\parbox{0.95\linewidth}{%
			\begin{small}
				
				$\textsf{}$ The mainly ideal functionality $\mathcal{F}$ of our scheme as well as the corresponding information leakages.
				\medskip
				\begin{enumerate}[(i)]
					
					\item $\mathcal{F}.\textsf{CNNFeaExtra}(\mathcal{I})$:
					\begin{itemize}[]
						\item \textbf{Functionality}. Two CS collaboratively extra the feature from the image $\mathcal{I}$, the shared feature is stored in each server.
						\item \textbf{Information leakage}. The information leaked here includes the image size of $\mathcal{I}$.
					\end{itemize}
					\medskip
					
					\item $\mathcal{F}.\textsf{SecPCA}(X)$:
					\begin{itemize}[]
						\item \textbf{Functionality}. Two CS collaboratively compress the matrix $X$, the shared of compressed matrix is stored in each server.
						\item \textbf{Information leakage}. The information leaked here includes the dim of the matrices, one similar matrix of $X$, and the square sum of the unnormalized eigenvectors.
					\end{itemize}
					
					\item $\mathcal{F}.\textsf{SecSort}(\{u\})$:
					\begin{itemize}[]
						\item \textbf{Functionality}. Two CS collaboratively sort the disordered distances $\{u\}$.
						\item \textbf{Information leakage}. The information leaked here includes the element number and size relationship of $\{u\}$, and the ratio between the difference of numbers in $\{u\}$.
					\end{itemize}
					
					\item $\mathcal{F}.\textsf{HKMIndex}(\{fea\})$:
					\begin{itemize}[]
						\item \textbf{Functionality}. Two CS collaboratively index the feature vectors $\{fea\}$ by HKM index algorithm.
						\item \textbf{Information leakage}. The information leaked here includes the element number of $\{fea\}$, the similarity relationship between the vectors and the corresponding images, and the ratio of the distances between corresponding images.
					\end{itemize}
					\medskip
					
					\item $\mathcal{F}.\textsf{HKMQuery}(Qfea, \{fea\})$:
					\begin{itemize}[]
						\item \textbf{Functionality}. Two CS collaboratively use the feature of query $Qfea$ to search the similar vectors in $\{fea\}$ by HKM query algorithm.
						\item \textbf{Information leakage}. The information leaked here includes the element number of $\{fea\}$, the similarity relationship between the query and the corresponding images, and the ratio of distances between query and corresponding images.
					\end{itemize}
					\medskip
					
					\item $\mathcal{F}.\textsf{C2LSHIndex}(\{fea\})$:
					\begin{itemize}[]
						\item \textbf{Functionality}. Two CS collaboratively index the feature vectors $\{fea\}$ by C2LSH index algorithm.
						\item \textbf{Information leakage}. The information leaked here includes the element number of $\{fea\}$, the similarity relationship between the vectors and the corresponding images, and the ratio of the distances between corresponding images.
					\end{itemize}
					\medskip
					
					\item $\mathcal{F}.\textsf{C2LSHQuery}(Qfea, \{fea\})$:
					\begin{itemize}[]
						\item \textbf{Functionality}. Two CS collaboratively use the feature of query $Qfea$ to search the similar vectors in $\{fea\}$ by C2LSH query algorithm.
						\item \textbf{Information leakage}. The information leaked here includes the element number of $\{fea\}$, the similarity relationship between the query and the corresponding images, and the ratio of distances between query and corresponding images.
					\end{itemize}
					\medskip
					
				\end{enumerate}
			\end{small}
		}
	}
	\caption{The functionality $\mathcal{F}$ and the information leakage in PPCBIR}
	\label{fig:FunctionLeakage}
\end{figure*}

\subsection{Remarks}\label{subsec:remarks}

In the above sections, for simplicity, the $secrets$ and $shares$ are assumed in $\mathbb{R}$; however, a real computer can only cope with the numbers in limited range and precision (i.e., a finite field $\mathbb{F}$). In detail, for the convenience of analysis on security and accuracy, the fixed-point numbers with $l_I$ bits on integer place and $l_D$ bits on the decimal place are considered in this subsection. Considering all the $shares$ and $secrets$ discussed above are represented in the $\mathbb{F}$, in this case, we analyze the influence brought by the real computer on accuracy and security in PPCBIR.

\subsubsection{Influence on accuracy}

To ensure the $shares$ in $\mathcal{F}$, the results of all multiplication computations should be truncated, in other words, the last $l_D$ bits on decimal place will be ignored.

Notably, an actual task is always robust for the small changes on inputs. With high probability, each addition of the additive $shares$ will bring at most one least significant bit error \cite{mohassel2017secureml}. Therefore, the main risk on accuracy lies in the underflow (i.e., close to 0) and overflow (i.e., close to $\infty$) problem caused by multiplication. 

Our scheme involves two kinds of multiplication: the $share$ and constant number; the $share$ and $share$. For the multiplication on $share$ and constant number, please note that the absolute value of the constant numbers (e.g., weights in the neural network) are, generally speaking, over $2^{-5}$. In this case, as long as the initial $shares$ split by $\mathcal{T}$ are not too small, the result of multiplication will not underflow. The risks on overflow problem are alike and we omit the analysis for simplicity.

For the multiplication on $share$ and $share$. We limit all the absolute values of the $shares$ in the range of $(2^{\delta-l_D}, 2^{\epsilon-l_D})\cup\{0\}$, where $\delta \geq \frac{l_D}{2}, \epsilon \leq \frac{l_I}{2}+l_D$. For simplicity, let us called the range $\mathbb{F}_s$. In this case, after one round of multiplication, the $share$ is still representable in $\mathcal{F}$. Notably, all protocols in our work will execute at most one round of multiplication on two $shares$ between the interactions.

Before the interaction, the server which owns the $share$ overflow $\mathbb{F}_s$ can execute the following operation: one server can seek a number in the $\mathbb{F}_s$ randomly, then sends the difference to the other server; the server which owns the $share$ underflow $\mathbb{F}_s$ can reset the $share$ as 0 directly in that the $share$ close to 0 is meaningless in the framework of additive secret sharing.

Besides, please note that the multiplications on $shares$ in PPCBIR are always used for comparison (e.g., \texttt{SecCmp}) or solving eigenvectors (e.g., algorithm \ref{alg: SecPCA}). In this case, the intermediate values (e.g., $f$ in \texttt{SecCmp}) will be discarded after getting the results. Therefore, the errors can not accumulate continuously to affect the actual task.

To sum up, with the help of modern computers, we can ensure the correctness and robustness of secure computation with the cost of some storage overhead and negligible precision loss.

\subsubsection{Influence on security}

The truncation operations are executed locally, thus it will not cause any security risk. Therefore, we focus on the risks brought by the recover of $share$ (e.g., $f$ in \texttt{SecCmp}).

Let us take $\texttt{SecCmp}$ as the example. To analyse the potential security risk in $\mathbb{F}$ accurately, we first transform the numbers in $\mathbb{F}$ to the numbers in $\mathbb{Z}$. In detail, considering $t$ and $r$ in $\texttt{SecCmp}$, we can get two integer $t'$ and $r'$ by letting $t' = 2^{l_D}t$, $r' = 2^{l_D}r$. Here, $r$ is the random number provided by $\mathcal{T}$, and $t$ is the value needed to be covered. Further, we define $z' = u'v' + 2^{l_D}k'$ and $z = uv+k$. Therefore, we can get the formula \ref{eq:PotentialLeak}, where $\eta$ represents a small ratio.

\begin{equation}
\label{eq:PotentialLeak}
\begin{array}{c}
\left\{
\begin{aligned}
& z\times{}2^{2\times{}l_D} \leq z' < ((z\cdot{}2^{l_D})+1)\times{2^{l_D}} \\
& z' = t'r' + 2^{l_D}k' \\
& 2^{\delta} \leq |r'| < 2^{\epsilon} \\
& |k'| < |\eta{}y'| \\
& 2^{\delta} \leq |k'| < 2^{\epsilon}
\end{aligned}
\right.

\end{array}
\end{equation}

For simplicity, $z$ is assumed to be positive. By formula \ref{eq:PotentialLeak}, we can get that $|t'|$ is a potential number in the range of $(z\times{}2^{2l_D-\epsilon}-\eta2^{l_D}, (z\times{}2^{l_D}+1)\times{}2^{l_D-\delta}+\eta2^{l_D})$, $|t'|\in{\mathbb{Z}}$. At the same time, $2^{\delta}<|t'|<2^{\epsilon}$. Thus, it will lead to the following three cases.

\emph{Case 1:} $z\leq{}2^{\delta+\epsilon-2l_D}-\eta2^{\delta-l_D}-\frac{1}{2^l_D}$. Then the $|t'|$ $\in$ $(2^{\delta}, (z\times{}2^{l_D}+1)\times{}2^{l_D-\delta}+\eta2^{l_D}]$, $|t'|$ $\in$ $\mathbb{Z}$.

\emph{Case 2:} $2^{\delta+\epsilon-2l_D}-\eta2^{\delta-l_D}-\frac{1}{2^l_D}<z<2^{\delta+\epsilon-2l_D}+\eta2^{\epsilon-l_D}$. Then the $|t'|$ $\in$ $(2^{\delta}, 2^{\epsilon})$, $|t'|$ $\in$ $\mathbb{Z}$.

\emph{Case 3:} $z\geq2^{\delta+\epsilon-2l_D}+\eta2^{\epsilon-l_D}$. Then the $|t'|$ $\in$ $[z\times{}2^{2l_D-\epsilon}-\eta2^{l_D}, 2^{\epsilon})$, $|t'|$ $\in$ $\mathbb{Z}$.

The above three cases imply that the information leakage when we use the finite field $\mathbb{F}$. However, as case 2 shows, generally speaking, the range is too broad to infer the detailed value of $x'$. The exception is that all the $t$, $r$, and $k$ achieve the maximum or minimum value; however, for an actual task, we believe the situation is negligible in the modern machine.

To sum up, the closer the $\mathbb{F}$ is to the $\mathbb{R}$, the less the leakage information is. Thus, with the help of modern machines, we believe that these protocols are feasible and secure in the PPCBIR task and many other actual outsourced tasks. Detailedly, in this work, the following settings are utilized in the experiments:

1) Using 24-byte floating numbers in python to represent the $share$ and $secret$. It is approximately equivalent to that $l_{D}=1021$, $l_{I}=1024$. As the $secret$ in PPCBIR is always a small number, we only assume $l_D = 64$, $l_I = 32, \delta = 32, \epsilon = 80$. In other words, the $secret$ and $share$ whose absolute values out the range of $(2^{-32}, 2^{16})$ will be reshared or marked as 0, even if it is supported by the computer. By this restriction, $\mathcal{T}$ only needs to generate the share in a small but enough interval. Please note that, for better accuracy, the truncation of $share$ will be executed only before the interaction.

2) After the initial conversion, the input numbers of VGG16 are in the range of $[0,1]$. Thus, $\mathcal{T}$ generate the $share$ whose absolute value in $[2^{-8},1]$ uniform randomly to cover the input. Besides, the original RGB value of the image will be covered by the matrix whose absolute values are the integers in the range of $[0,256]$. For the efficiency of decryption, both the four $shares$ will be sent to the servers.

3) $\mathcal{T}$ chooses the sign of $share$ (e.g., $a_1$, $t_1$) or $secret$ (e.g., $a$, $t$) uniform randomly. The $share$ is generated uniformly randomly in $\mathbb{F}_s$. To decrease the risk of overflow, the exponential value of $secret$ is generated with the normal distribution, then the detailed $secret$ is uniformly generated in the corresponding range. The $\eta$ is set as $2^{-8}$.

\section{Experiment results}\label{sec:experimentRes}

The section evaluates the performance of the proposed scheme in terms of encryption effectiveness, retrieval accuracy and retrieval efficiency. We implement the proposed scheme  on Windows 10 operating system. All the user side experiments (i.e., image owner, authorized user, and trusty third party) are executed in a machine with Intel Core i5-8250u CPU @ 1.6GHZ and 16 GB memory. The cloud side experiments are executed on two servers running Windows 10 with the LAN setting. Each server is equipped with Intel Core i7-9700 CPU @ 3GHZ and 16 GB memory. We use Corel-1k and Corel-10k image dataset as the experimental dataset. The images in Corel-1k size either 384$\times$256 or 256$\times$384. Corel-1k includes 10 categories of images and each category contains 100 similar images. The size of images in Corel-10k is either 187$\times$126 or 126$\times$187. Corel-10k includes 100 categories of images and each category contains 100 similar images.

As described in subsection \ref{subsec:PPCBIR}, the schemes in the first category let the image owner undertake the feature extraction task, which is obviously inferior to our scheme. Therefore, we focus on the comparison of schemes in the second category. We first give the comparison of the methods which execute the privacy-preserving CNN inference.

\subsection{Performance of Privacy-Preserving CNN Inference Protocol}

Due to the advantage of our schemes benefit from the designed protocols, we first show the comparison results of the complexity of protocols which are shown in Table \ref{tab:ProtocolComp}. \cite{huang2019lightweight} compares the sign of numbers based on the most significant bit, therefore, their protocol needs $O(l)$ rounds interaction, here $l$ means the bit-length of ciphertext, which is generally set as 32. \cite{xiong2020efficient} gives their scheme based on additive resharing and multiplicative resharing, however, the transformation between them always leads to three or more interaction rounds.

\begin{table*}[ht]
	\centering
	\caption{Comparison of the protocol complexities\\(here $l$ means bit-width, and $n$ means the dimension of squared matrix)}
	\label{tab:ProtocolComp}
		\begin{tabular}{|l|c|c|c|c|c|c|}
			\hline
			\multirow{2}{*}{} & \multicolumn{2}{c|}{\texttt{SecCmp}} & \multicolumn{2}{c|}{\texttt{SecDiv}} & \multicolumn{2}{c|}{\texttt{SecMatInv}} \\ \cline{2-7} 
			& Rounds & Comm(bits) & Rounds & Comm(bits) & Rounds & Comm(bits) \\ \hline
			Huang\cite{huang2019lightweight} & $l+1$      & $10l-2$      & -      & -          & -      & -          \\ \hline
			Xiong\cite{xiong2020efficient} & $3$      & $2l+2$       &  3    & 6$l$         & -      & -          \\ \hline
			Ours   & 2      & 6$l$         & 2      & 10$l$         & 2      & 6$n^2l$    \\ \hline
		\end{tabular}
\end{table*}

To show the advantage in the scene of privacy-preserving CNN better, the real-time comparison is given in Table \ref{tab:FeedForwardConsumption}. Same as previous works\cite{juvekar2018gazelle}, the MNIST dataset which contains plenty of 28$\times$28 gray-scale images is used. The structure of used CNN model comes from the previous work \cite{liu2017oblivious}, which is quite similar to VGG16: 2-Conv and 2-FC with ReLU and MaxPool.

\begin{table}[]
	\centering
	\caption{Inference consumption in simple CNN model}
	\label{tab:FeedForwardConsumption}
	\begin{tabular}{|l|c|c|c|c|}
		\hline
		\multirow{2}{*}{} & \multicolumn{2}{c|}{Runtime(s)} & \multicolumn{2}{c|}{Message sizes(MB)} \\ \cline{2-5} 
		& Offline & Online & Offline & Online \\ \hline
		MiniONN\cite{liu2017oblivious} & 472     & 72     & 3046    & 6226   \\ \hline
		Gazelle\cite{juvekar2018gazelle} & 9.34    & 3.56   & 940     & 296    \\ \hline
		Huang\cite{huang2019lightweight}   & 0.09    & 0.21   & 1.57    & 0.99   \\ \hline
		Ours    & 0.002   & 0.05   & 1.41    & 0.24   \\ \hline
	\end{tabular}
\end{table}

Since our scheme outsources a part of random number generation tasks to CS, the offline consumption also significantly decreases. As a baseline, the actual efficiency of generating random numbers or matrices is shown as follow: In one second, $\mathcal{T}$ can generate $1.7\times 10^7$ random share of $Beaver's$ $triples$; or $5.4\times 10^5$ random share of matrices $(A_i, B_i, (A\cdot B)_i)$, where $A_i$ sizes $1\times 8$ and $B_i$ sizes $8\times 1$; or $6.4\times 10^3$ random share of matrices $(A_i, B_i, (A\cdot B)_i)$, where $A_i$ sizes $512\times 8$ and $B_i$ sizes $8\times 1$. The baseline will be used in the subsection \ref{subsec:retrievalconsum}.

The above results show the advantage of our schemes compared to previous CNN inference protocols in time or storage consumption. Besides, our scheme does not need scaling to ensure that the protocol works correctly, accordingly, no loss in accuracy will be caused in theory. Therefore, to our knowledge, the most efficient and accurate results in PPCBIR schemes based on typical features are introduced in this work. In the following, we only compare our work to the schemes based on the statistic feature.

\subsection{Consumption before retrieval}
In this section, we focus on the process from the step that image owner encrypts his images to the step that the cloud servers finish the index building. Two typical schemes \cite{xia2019boew,cheng2016encrypted} based on statistics are chosen as the comparative experiments, we also give the results in plaintext state as the baseline and comparison.

\subsubsection{Image outsourcing}

As described in subsection \ref{subsec:remarks}, in our scheme, the image owner only needs to split the image and corresponding input into two different $shares$. The schemes based on statistics need permutations on the image, which leads to plenty of time consumption. The time consumption of outsourcing whole Corel-1k or Corel-10k is shown in Table \ref{tab:ImageOutsource}.

\begin{table}[ht]
	\centering
	\caption{Time consumption of image outsourcing}
	\label{tab:ImageOutsource}
	\begin{tabular}{|l|c|c|c|}
		\hline
		& BOEW\cite{xia2019boew}                           & Cheng\cite{cheng2016encrypted}                       & Ours                        \\ \hline
		Corel-1k                        & 281.86s                       & 79.07s                      & 39.82s                      \\ \hline
		Corel-10k 						& 1118.33s 			& 407.7s 			& 202.71s \\ \hline
	\end{tabular}
\end{table}

\subsubsection{Feature extraction and aggregation}

As described in subsection \ref{subsec:FeaExtra}, the encrypted feature will be collaboratively extracted by two servers. Due to plenty of non-linear operations, the consumption exceeds the schemes based on statistics. However, on the one hand, the aggregation consumption is much lower in our scheme; on the other hand, the time consumption of extraction can be reduced if simpler CNN models are used. The total time consumption of the two datasets is shown in Table \ref{tab:FeatureExtra}.

\begin{table*}[ht]
	\centering
	\caption{Time consumption of feature extraction and aggregation}
	\label{tab:FeatureExtra}
	\begin{tabular}{|l|c|c|c|c|c|c|c|c|}
		\hline
		\multirow{2}{*}{} & \multicolumn{2}{c|}{BOEW\cite{xia2019boew}} & \multicolumn{2}{c|}{Cheng\cite{cheng2016encrypted}} & \multicolumn{2}{c|}{Ours} & \multicolumn{2}{c|}{Plaintext} \\ \cline{2-9} 
		& Corel-1k & Corel-10k & Corel-1k & Corel-10k & Corel-1k & Corel-10k & Corel-1k & Corel-10k \\ \hline
		Feature extraction  & 190.99s  & 1495.07s  & 109.79s  & 426.33s   & 4015.61s & 10365.71s  & 226.34s  & 468.32s   \\ \hline
		Feature aggregation & 620.81s  & 860.73s   & -        & -         & 0.028s   & 0.279s    & 0.028s   & 0.279s    \\ \hline
	\end{tabular}
\end{table*}

The communication size during feature extraction is given in Table \ref{tab:FeatureExtraComm}. The tasks in this subsection are all executed in the \emph{offline phase}, which means $\mathcal{T}$ has sufficient time for the generation of random numbers or matrices. Therefore, in this subsection, only the communication consumption during executing the task is given.

\begin{table}[ht]
	\centering
	\caption{Communication size of feature extraction}
	\label{tab:FeatureExtraComm}
	\begin{tabular}{|l|c|l|l|l|c|l|l|l|}
		\hline
		\multicolumn{1}{|c|}{\multirow{2}{*}{}} & \multicolumn{4}{c|}{\multirow{2}{*}{Corel-1k}} & \multicolumn{4}{c|}{\multirow{2}{*}{Corel-10k}} \\
		\multicolumn{1}{|c|}{} & \multicolumn{4}{c|}{}         & \multicolumn{4}{c|}{}          \\ \hline
		Feature extraction     & \multicolumn{4}{c|}{937.32GB} & \multicolumn{4}{c|}{2214.74GB} \\ \hline
	\end{tabular}
\end{table}

\subsubsection{Feature Compression}

In this work, the dimension of the compressed feature is set as $8$ for both Corel-1k and Corel-10k. Since the previous schemes in PPCBIR ignore the process of compression, we only report the experiment results of our scheme which are shown in Table \ref{tab:FeaCompress}. It can be seen that the consumption in compression is negligible compared to the above step, however, we will show that it will highly decrease the communication consumption in the following.

\begin{table}[ht]
	\centering
	\caption{Consumption of feature compression}
	\label{tab:FeaCompress}
	
	\begin{tabular}{|l|c|l|c|l|c|l|l|l|}
		\hline
		\multirow{3}{*}{} & \multicolumn{4}{c|}{Ours}                                   & \multicolumn{4}{c|}{Plaintext} \\ \cline{2-9} 
		& \multicolumn{2}{c|}{\multirow{2}{*}{Runtime}} & \multicolumn{2}{c|}{\multirow{2}{*}{Message size}} & \multicolumn{4}{c|}{\multirow{2}{*}{Runtime}} \\
		& \multicolumn{2}{c|}{}       & \multicolumn{2}{c|}{}         & \multicolumn{4}{c|}{}          \\ \hline
		Corel-1k          & \multicolumn{2}{c|}{1.719s} & \multicolumn{2}{c|}{31.85MB}  & \multicolumn{4}{c|}{0.297s}    \\ \hline
		Corel-10k         & \multicolumn{2}{c|}{3.592s} & \multicolumn{2}{c|}{137.31MB} & \multicolumn{4}{c|}{0.375s}    \\ \hline
	\end{tabular}
\end{table}

\subsubsection{Index Building}

In our experiments, the hyper-parameter $k$ in HKM is set as $4$, and the minimum number limitation of vectors in $candidates$ is set as the $3\times m$ in both HKM and C2LSH, where $m$ is the number of returned images (e.g., 50). The other parameters of C2LSH are set as suggested in \cite{gan2012locality}.

Table \ref{tab:IdxBuild} gives the time consumption of two index building schemes described in subsection \ref{subsec:IndexBuild}. Here the results on both uncompressed and compressed features are given, and it is easy to notice that the communication consumption gain significantly decreases. For example, the communication consumption during HKM building on the compressed feature is only about $1/38$ of the uncompressed version in Corel-1k.

\begin{table*}[]
	\centering
	\caption{Consumption of Index Building}
	\label{tab:IdxBuild}
	\begin{tabular}{|c|c|l|l|l|l|l|c|c|c|c|c|c|}
		\hline
		\multicolumn{7}{|c|}{\multirow{3}{*}{}}               & \multicolumn{3}{c|}{Corel-1k}         & \multicolumn{3}{c|}{Corel-10k}        \\ \cline{8-13} 
		\multicolumn{7}{|c|}{}                                & \multicolumn{2}{c|}{Ours} & Plaintext & \multicolumn{2}{c|}{Ours} & Plaintext \\ \cline{8-13} 
		\multicolumn{7}{|c|}{}                                & Runtime  & Message sizes  & Runtime   & Runtime  & Message sizes  & Runtime   \\ \hline
		\multirow{2}{*}{HKM}   & \multicolumn{6}{c|}{512-dim} & 10.13s   & 781.69MB       & 1.05s     & 275.63s  & 25.74GB        & 12.97s    \\ \cline{2-13} 
		& \multicolumn{6}{c|}{8-dim}   & 2.44s    & 20.87MB        & 0.75s     & 46.67s   & 569.19MB       & 7.72s     \\ \hline
		\multirow{2}{*}{C2LSH} & \multicolumn{6}{c|}{512-dim} & 1.43s    & 6.81MB         & 0.52s     & 15.99s   & 65.61MB         & 6.18s     \\ \cline{2-13}
		& \multicolumn{6}{c|}{8-dim}   & 1.17s    & 1.99MB         & 0.39s     & 12.98s    & 25.88MB        & 5.23s     \\ \hline
	\end{tabular}
\end{table*}

\subsection{Retrieval Consumption and Precision}

\subsubsection{Retrieval Consumption}\label{subsec:retrievalconsum}

Before the authorized user gets the plaintext images he needs, there are mainly three steps as described in subsection \ref{subsec:SecRetrieval}: $\emph{Trapdoor generation}$, $\emph{similarity computation}$ $\emph{in CS}$, and $\emph{decryption}$.

The time consumption comparison results on retrieval (return Top-50 similar images) from Corel-1k and Corel-10k are shown in Table \ref{tab:RetrievalCorel-1k} and Table \ref{tab:RetrievalCorel-10k}. From the tables, it can be seen that although the feature extraction of our scheme needs more time compared to statistics based schemes. However, the advantage of encryption and decryption will make our total consumption smaller. The non-index (i.e., linear index) situation is also given for the comparison. It is easy to notice that the utilization of index greatly decreases communication costs, and makes all kinds of consumption not substantially increase as the size of the dataset increase.

\begin{table*}[ht]
	\centering
	\caption{Retrieval Consumption in Corel-1k (Top-50)}
	\label{tab:RetrievalCorel-1k}
	\begin{tabular}{|l|l|l|l|l|l|l|c|c|c|c|c|c|}
		\hline
		\multicolumn{7}{|c|}{\multirow{3}{*}{}} &
		\multirow{2}{*}{BOEW\cite{xia2019boew}} &
		\multirow{2}{*}{Cheng\cite{cheng2016encrypted}} &
		\multicolumn{3}{c|}{Ours} &
		\multirow{2}{*}{Plaintext} \\ \cline{10-12}
		\multicolumn{7}{|c|}{} &
		&
		&
		Offline &
		\multicolumn{2}{c|}{Online} &
		\\ \cline{8-13} 
		\multicolumn{7}{|c|}{} &
		Runtime &
		Runtime &
		Runtime &
		Runtime &
		Message sizes &
		Runtime \\ \hline
		\multicolumn{7}{|l|}{Trapdoor generation} &
		0.28s &
		0.08s &
		- &
		0.024s &
		- &
		- \\ \hline
		\multirow{6}{*}{\begin{tabular}[c]{@{}l@{}}Similarity \\ computation \\ in CS\end{tabular}} &
		\multicolumn{6}{l|}{feature extraction} &
		0.19s &
		0.11s &
		5.106s &
		4.01s &
		959.76MB &
		0.226s \\ \cline{2-13} 
		&
		\multicolumn{6}{l|}{feature aggregation} &
		0.01s &
		- &
		- &
		\textless{}1ms &
		- &
		\textless{}1ms \\ \cline{2-13} 
		&
		\multicolumn{6}{l|}{feature compression} &
		- &
		- &
		0.2ms &
		0.14s &
		36.22KB &
		\textless{}1ms \\ \cline{2-13} 
		&
		\multicolumn{3}{l|}{\multirow{1}{*}{retrieval}} &
						\multicolumn{3}{l|}{HKM} &
						- &
						- &
						1.088ms &
						0.182s &
						51.25KB &
						\textless{}1ms \\ \cline{5-13} 
						&
						\multicolumn{3}{l|}{} &
						\multicolumn{3}{l|}{C2LSH} &
						- &
						- &
						1.27ms &
						0.209s &
						38.47KB &
						0.001s \\ \cline{5-13} 
						&
						\multicolumn{3}{l|}{} &
		\multicolumn{3}{l|}{Linear} &
		0.003s&
		54.35s&
		3.17ms &
		0.199s &
		141.06KB &
		0.003s \\ \hline
		\multicolumn{7}{|l|}{Decryption} &
		14.09s &
		3.92s &
		- &
		0.85s &
		- &
		- \\ \hline
		\multicolumn{7}{|l|}{Total} &
		14.575s&
		58.52s&
		\begin{tabular}[c]{@{}c@{}}5.107s/\\ 5.107s/\\ 5.109s\end{tabular} &
		\begin{tabular}[c]{@{}c@{}}5.189s/\\ 5.216s/\\ 5.206s\end{tabular} &
		\begin{tabular}[c]{@{}c@{}}959.85MB/\\ 959.83MB/\\ 959.93MB\end{tabular} &
		\begin{tabular}[c]{@{}c@{}}0.226s/\\ 0.227s/\\ 0.229s\end{tabular} \\ \hline

	\end{tabular}
\end{table*}

\begin{table*}[ht]
	\centering
	\caption{Retrieval Consumption in Corel-10k (Top-50)}
	\label{tab:RetrievalCorel-10k}
	\begin{tabular}{|l|l|l|l|l|l|l|c|c|c|c|c|c|}
		\hline
		\multicolumn{7}{|c|}{\multirow{3}{*}{}} &
		\multirow{2}{*}{BOEW\cite{xia2019boew}} &
		\multirow{2}{*}{Cheng\cite{cheng2016encrypted}} &
		\multicolumn{3}{c|}{Ours} &
		\multirow{2}{*}{Plaintext} \\ \cline{10-12}
		\multicolumn{7}{|c|}{} &
		&
		&
		Offline &
		\multicolumn{2}{c|}{Online} &
		\\ \cline{8-13} 
		\multicolumn{7}{|c|}{} &
		Runtime &
		Runtime &
		Runtime &
		Runtime &
		Message sizes &
		Runtime \\ \hline
		\multicolumn{7}{|l|}{Trapdoor generation} &
		0.112s &
		0.04s &
		- &
		0.011s &
		- &
		- \\ \hline
		\multirow{6}{*}{\begin{tabular}[c]{@{}l@{}}Similarity \\ computation \\ in CS\end{tabular}} &
		\multicolumn{6}{l|}{feature extraction} &
		0.149s &
		0.07s&
		1.182s &
		1.05s &
		226.83MB &
		0.046s \\ \cline{2-13} 
		&
		\multicolumn{6}{l|}{feature aggregation} &
		0.01s&
		- &
		- &
		\textless{}1ms &
		- &
		\textless{}1ms \\ \cline{2-13} 
		&
		\multicolumn{6}{l|}{feature compression} &
		- &
		- &
		0.2ms &
		0.14s &
		36.22KB &
		\textless{}1ms \\ \cline{2-13} 
		&
				\multicolumn{3}{l|}{\multirow{3}{*}{retrieval}} &
				\multicolumn{3}{l|}{HKM} &
				- &
				- &
				1.83ms &
				0.196s &
				98.16KB &
				\textless{}1ms \\ \cline{5-13} 
				&
				\multicolumn{3}{l|}{} &
				\multicolumn{3}{l|}{C2LSH} &
				- &
				- &
				1.52ms &
				0.211s &
				53.49KB &
				0.001s \\ \cline{5-13} 
				&
				\multicolumn{3}{l|}{} &
				\multicolumn{3}{l|}{Linear} &
		0.03s&
		32.05s&
		30.87ms &
		0.261s &
		1406.68KB &
		0.029s \\ \hline
		\multicolumn{7}{|l|}{Decryption} &
		5.59s &
		2.03s &
		- &
		0.422s &
		- &
		- \\ \hline
		\multicolumn{7}{|l|}{Total} &
		5.991s&
		34.19s&
		\begin{tabular}[c]{@{}c@{}}1.184s/\\ 1.184s/\\ 1.213s\end{tabular} &
		\begin{tabular}[c]{@{}c@{}}1.811s/\\ 1.826s/\\ 1.876s\end{tabular} &
		\begin{tabular}[c]{@{}c@{}}226.96MB/\\ 226.92MB/\\ 228.24MB\end{tabular} &
		\begin{tabular}[c]{@{}c@{}}0.046s/\\ 0.047s/\\ 0.075s\end{tabular} \\ \hline
	\end{tabular}
\end{table*}

\subsubsection{Discussion on trusty third party}\label{subsec:Discussion on TTP}

From the above experiments, it is easy to see that only two types of matrices needed to be pre-generated for retrieval. The size of the matrix is determined by the dimension of the uncompressed and compressed features, which can be seen as hyper-parameters shared by all entities before retrieval. Therefore, $\mathcal{T}$ can easily generate enough random matrices before retrieval.

After giving the actual time consumption of retrieval, we briefly analyze the choice of trusty third party $\mathcal{T}$. It is fairly good that the government or other credible agencies which can be trusty to all participants are willing to undertake this role, however, it is often difficult to seek in the real world. In this case, a reasonable assumption is letting the owner of images play the role instead.

In detail, besides outsourcing images, the image owner also spends a small amount of computing resources to generate random numbers and matrices, and the operations during \emph{offline phase} are based upon them. Before the query, the authorized user also costs some time for generation, and servers will utilize them during the feature extraction of the query and secure compression process.

When facing actual distance computation, it is inevitable that the features of both image owner and authorized user will be involved. As the feature comes from authorized user is always meaningless to image owner, it is more suitable for image owner to undertake the generation task for the process. Since the random data needed during distance computation is about only 1ms as shown in Table \ref{tab:RetrievalCorel-1k} and \ref{tab:RetrievalCorel-10k}, which means about one thousand queries can be supported at the cost of only one second during the \emph{offline phase}. To sum up, we believe that the owner of images is suitable and capable of playing the role of $\mathcal{T}$.

\subsubsection{Retrieval accuracy}

\begin{figure}[htbp]
	\centering
	\subfigure[Comparison in Corel-1k]{
		\begin{minipage}[t]{0.45\linewidth}
			\centering
			\includegraphics[width=1.6in]{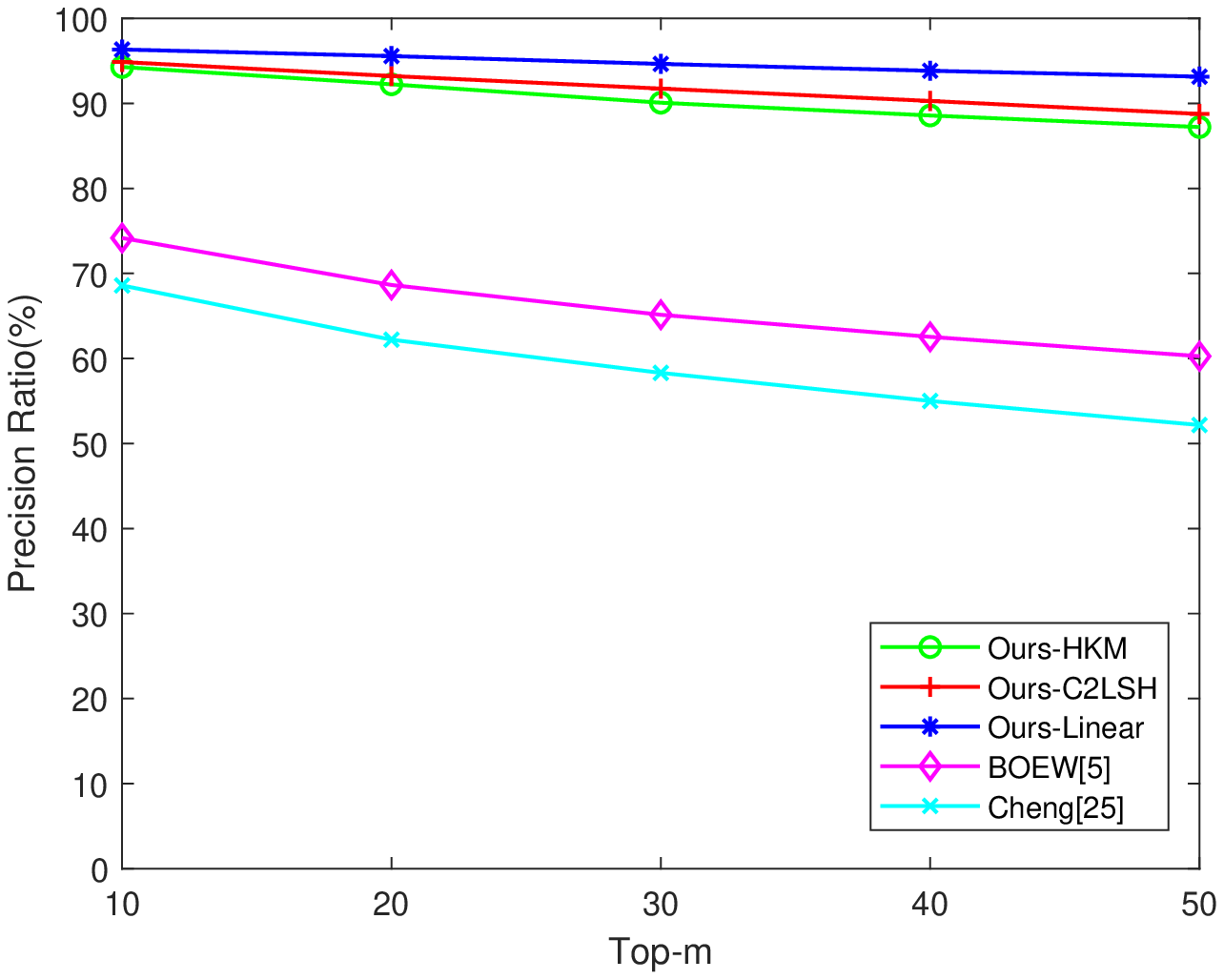}
		\end{minipage}%
	}
	\hspace{.1in}
	\subfigure[Comparison in Corel-10k]{
		\begin{minipage}[t]{0.45\linewidth}
			\centering
			\includegraphics[width=1.6in]{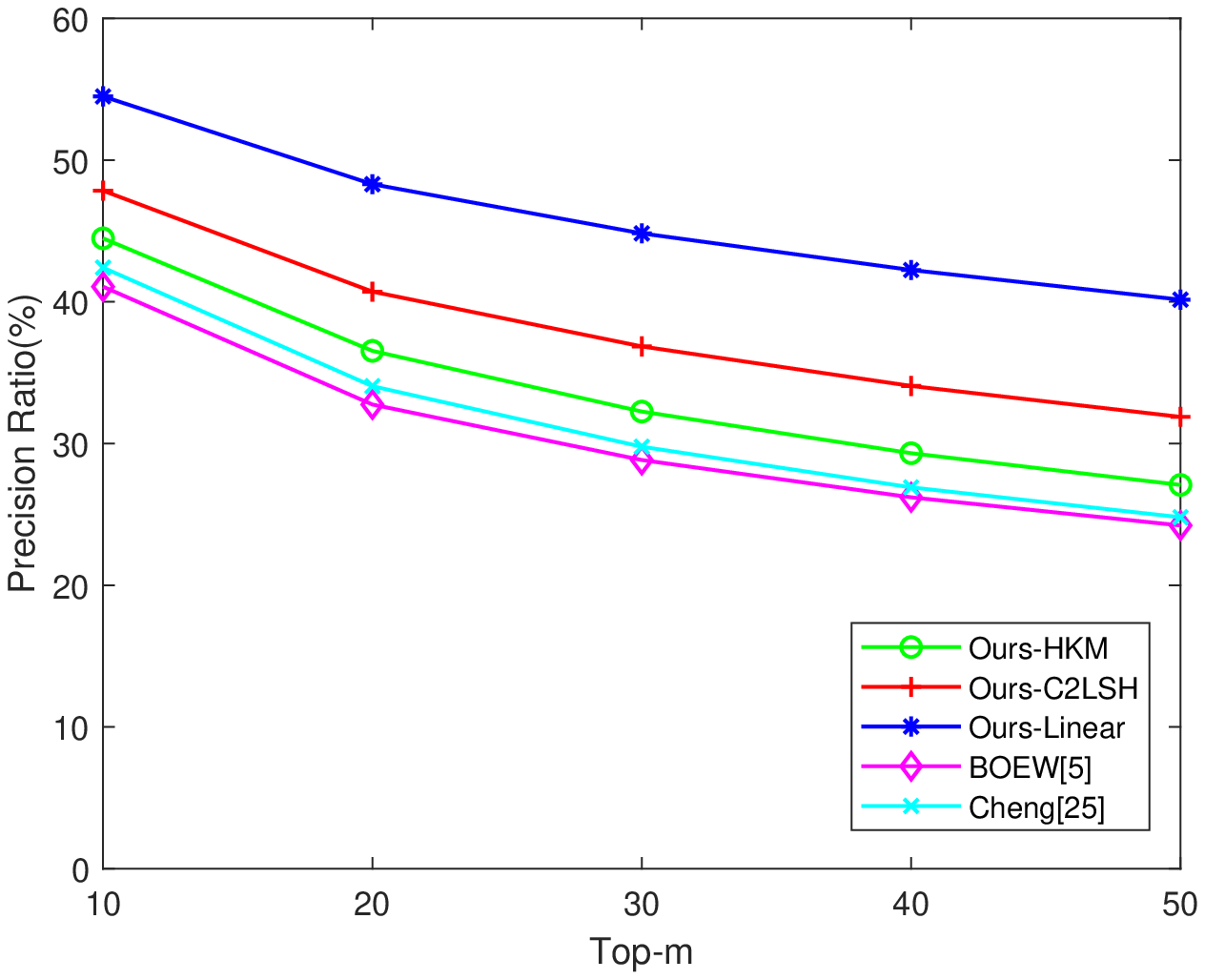}
		\end{minipage}%
	}
	\qquad
	\hspace{.1in}
	\subfigure[Comparison between compressed and uncompressed features in Corel-1k]{
		\begin{minipage}[t]{0.45\linewidth}
			\centering
			\includegraphics[width=1.6in]{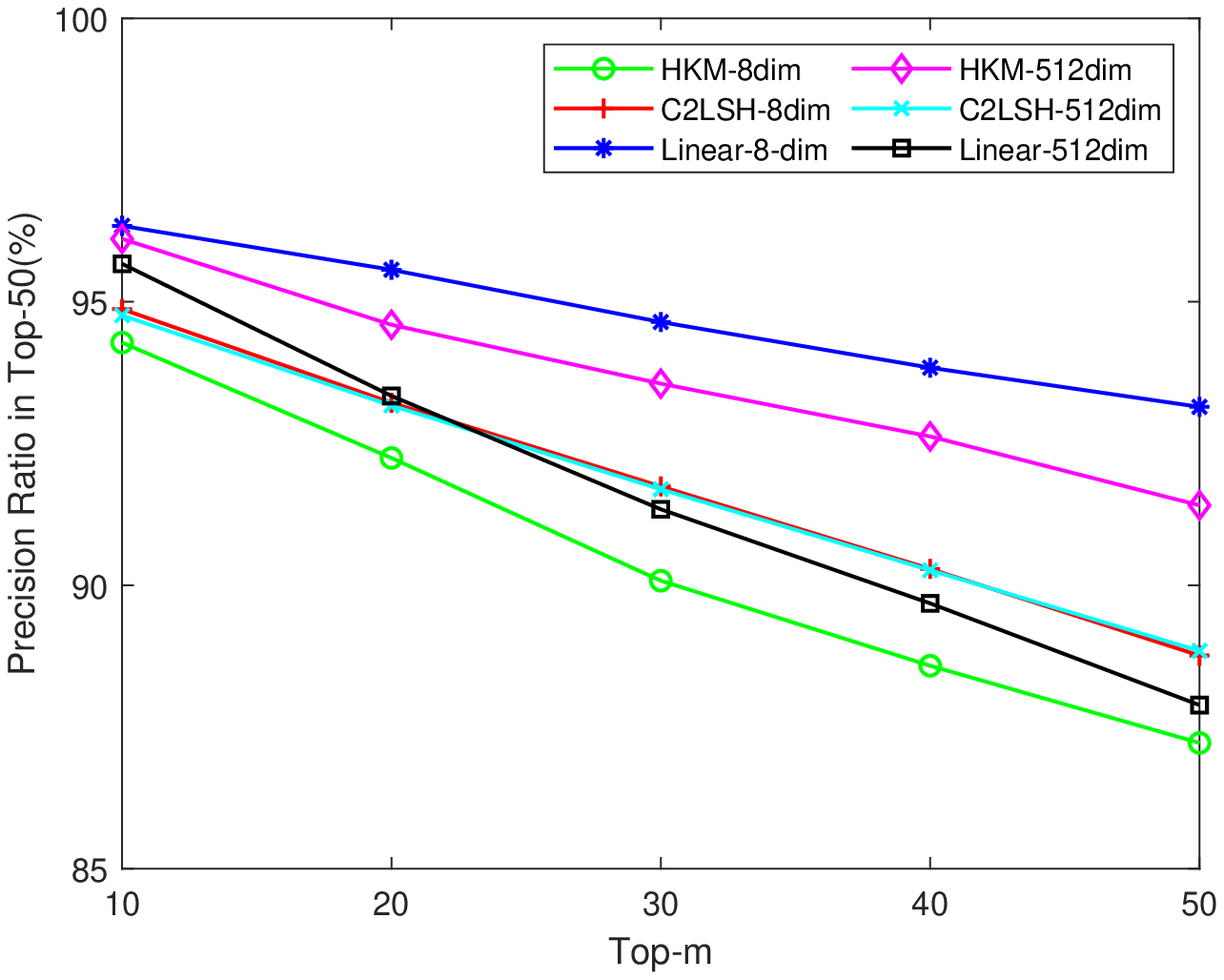}
		\end{minipage}
	}
	\hspace{.1in}
	\subfigure[Comparison between compressed and uncompressed features in Corel-10k]{
		\begin{minipage}[t]{0.45\linewidth}
			\centering
			\includegraphics[width=1.6in]{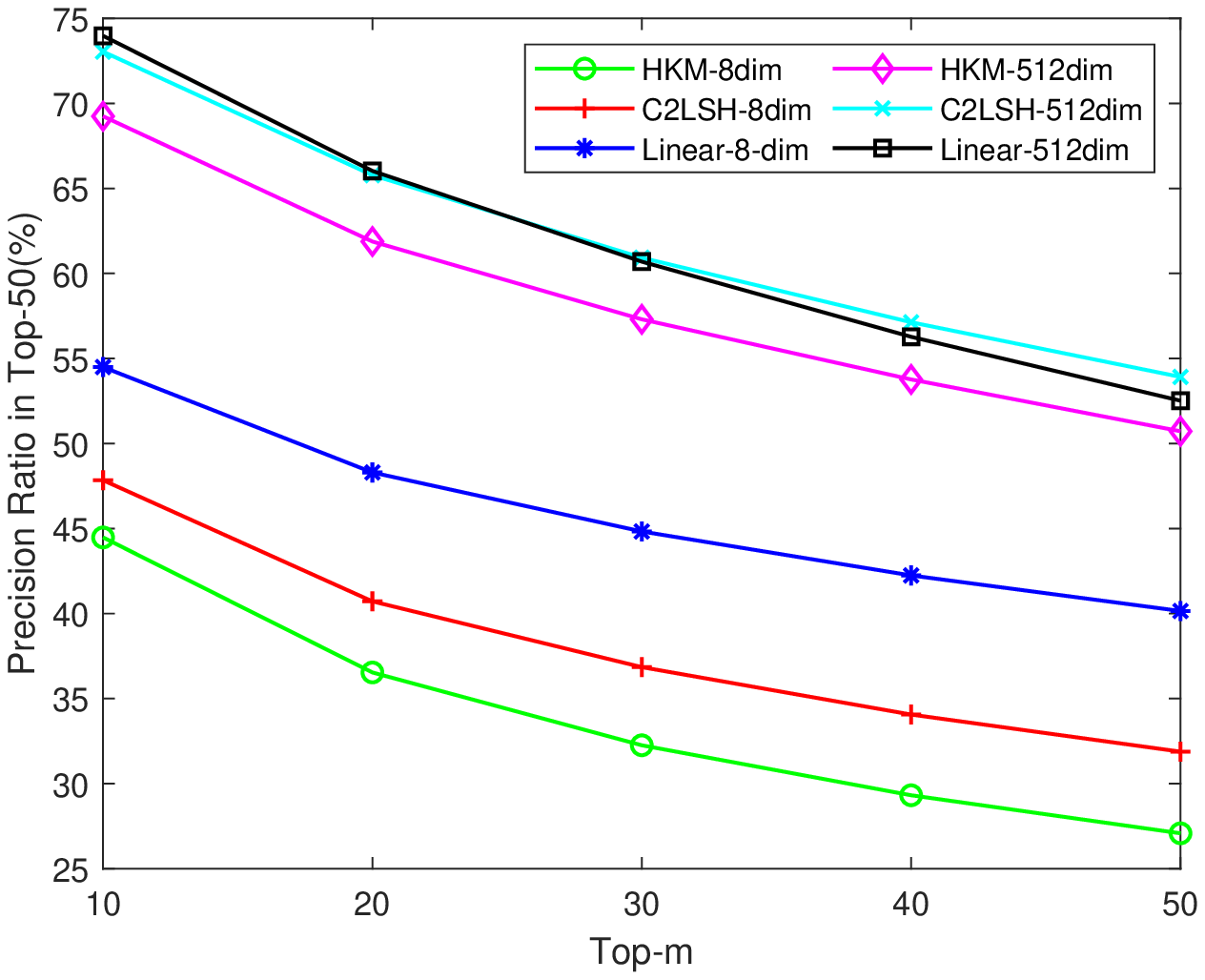}
		\end{minipage}
	}
	
	\centering
	\caption{Retrieval accuracy comparison in Corel-1k and Corel-10k}
	\label{fig:RetrievalAccuracyComparison}
\end{figure}

In our experiments, the "precision" of a query is defined as that in \cite{muller2001performance}: $P_m = m'/m$, where $m'$ is the number of real similar images in the $m$ retrieved images. We choose all the images in Corel-1k and Corel-10k, and the retrieval accuracy comparison in these two datasets is shown as Fig. \ref{fig:RetrievalAccuracyComparison}.

From the first and second sub-figures, due to the better feature extractor, it can be seen that our scheme shows great advantage when compared with the schemes based on statistics. As the third and fourth sub-figures show, we may observe that high dimension vectors show better accuracy. However, when the number of images is small (e.g., Corel-1k), the compressed feature shows better accuracy and the utilization of index (e.g., HKM) can even improve the accuracy by excluding the wrong candidate vectors brought by the feature extractor. Furthermore, the accuracy of our scheme is same as the corresponding plaintext state, in other words, the lossless property is confirmed by both theory and experiments.

\section{Conclusion}\label{sec:conclu}

In this paper, to bridge the gap between CBIR and PPCBIR, we utilize typical additive secret sharing and propose a series of novel protocols to execute secure computation efficiently. We further simulate the inference process of VGG16, the typical compression and index building schemes for better accuracy and efficiency. In future work, we will consider the construction of better protocols and other applications based on additive secret sharing.

\section*{Acknowledgements}
This work is supported in part by the Jiangsu Basic Research Programs-Natural Science Foundation under grant numbers BK20181407, in part by the National Natural Science Foundation of China under grant numbers U1936118, 61672294, in part by Six peak talent project of Jiangsu Province (R2016L13), Qinglan Project of Jiangsu Province, and “333” project of Jiangsu Province, in part by the National Natural Science Foundation of China under grant numbers U1836208, 61702276, 61772283, 61602253, and 61601236, in part by National Key R\&D Program of China under grant 2018YFB1003205, in part by the Priority Academic Program Development of Jiangsu Higher Education Institutions (PAPD) fund, in part by the Collaborative Innovation Center of Atmospheric Environment and Equipment Technology (CICAEET) fund, China. Zhihua Xia is supported by BK21+ program from the Ministry of Education of Korea.

\bibliographystyle{IEEEtran}

\bibliography{PPIRBASS}

\begin{IEEEbiography}[{\includegraphics[width=1in,height=1.25in,clip,keepaspectratio]{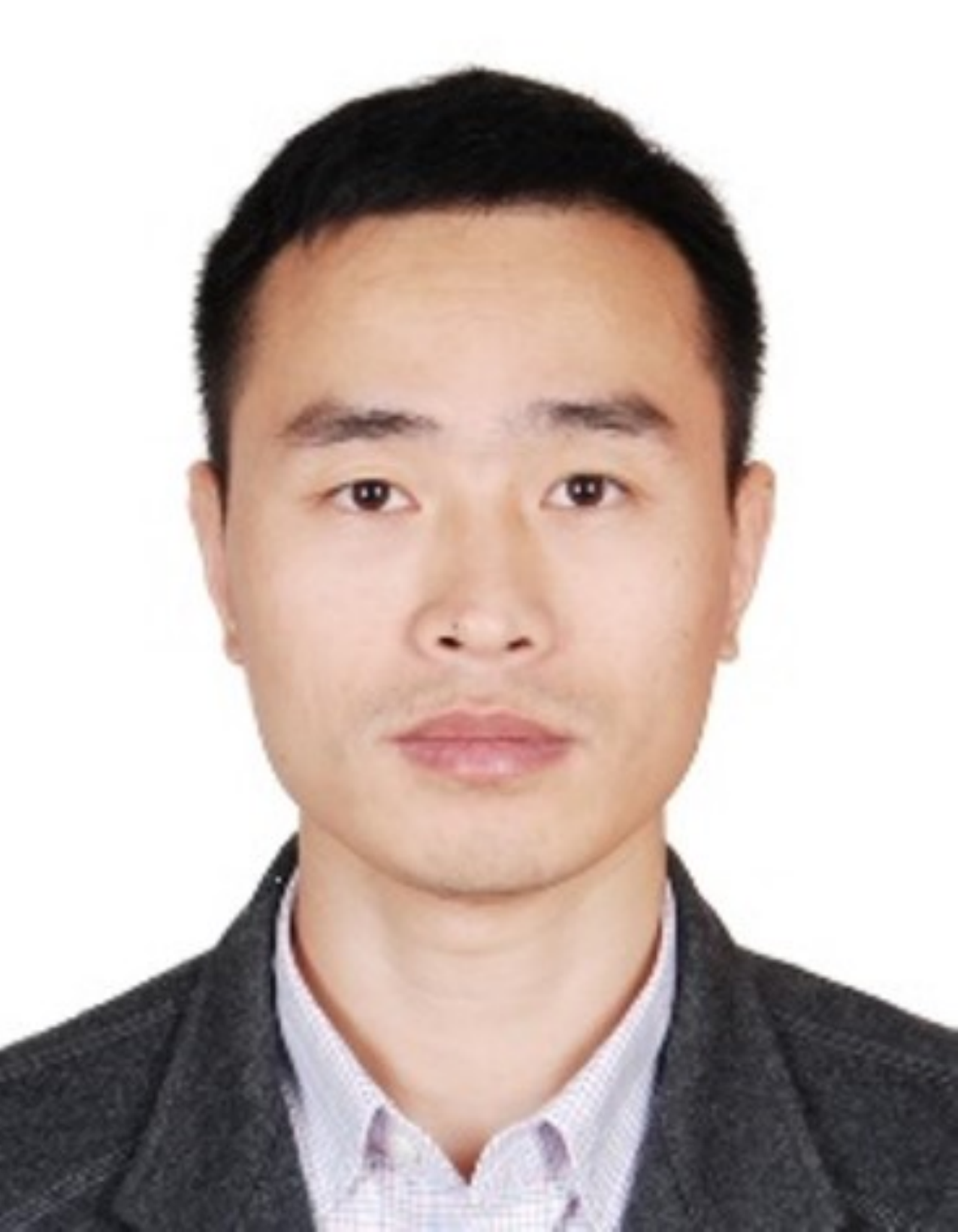}}]{Zhihua Xia }
	received his B.S. degree in Hunan City University, China, in 2006, and the Ph.D. degree in computer science and technology from Hunan University, China, in 2011.He is currently an associate professor with the School of Computer and Software, Nanjing University of Information Science and Technology, China. He was a visiting professor with the Sungkyunkwan University, Korea, 2016. His research interests include cloud computing security and digital forensic. He is a member of the IEEE.
\end{IEEEbiography}

\begin{IEEEbiography}[{\includegraphics[width=1in,height=1.25in,clip, keepaspectratio]{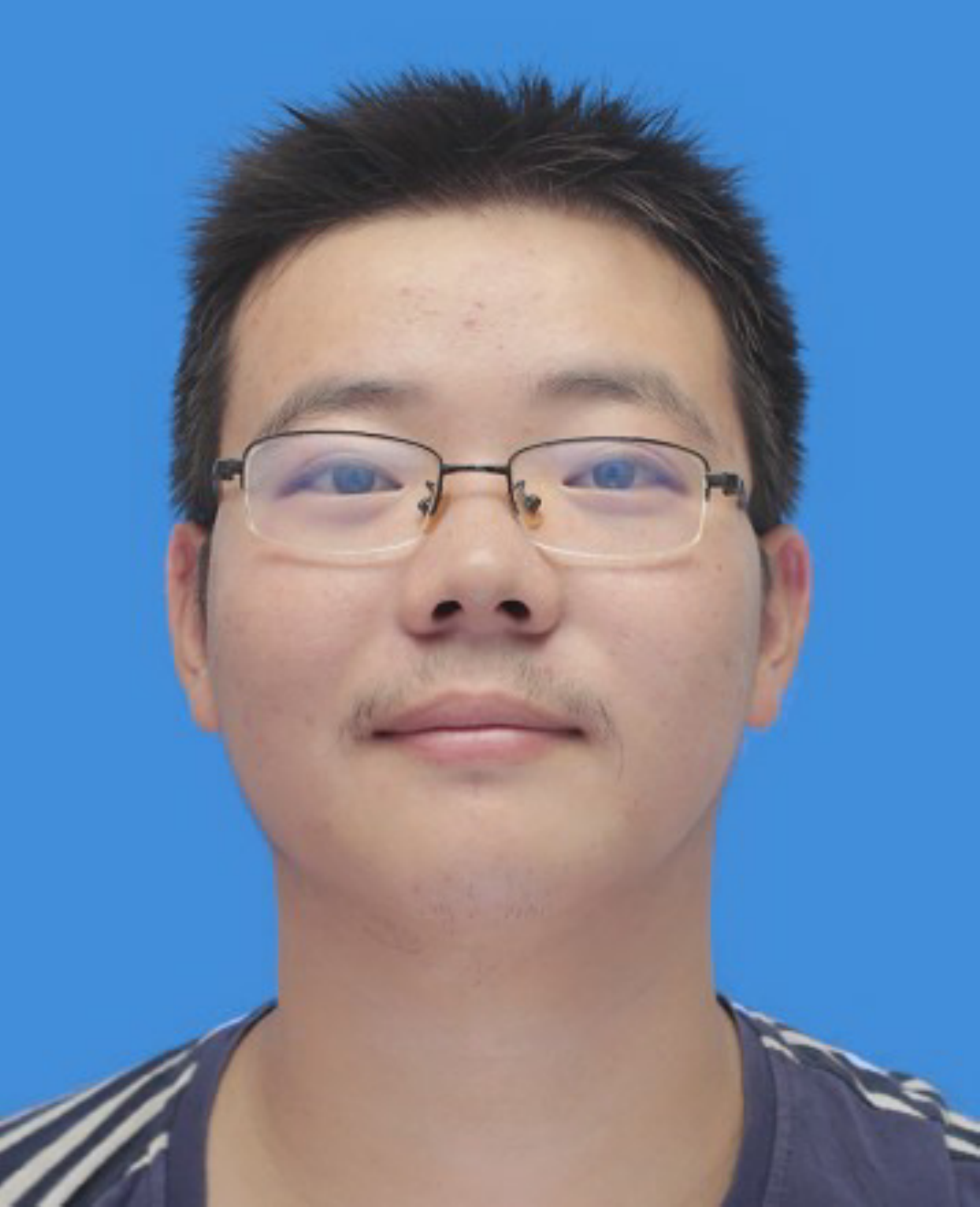}}]{Qi Gu}
	is currently pursing his master degree in the School of Computer and Software, Nanjing University of Information Science and Technology, China. His research interests include functional encryption, image retrieval and nearest neighbor search.
\end{IEEEbiography}

\begin{IEEEbiography}[{\includegraphics[width=1in,height=1.25in,clip,keepaspectratio]{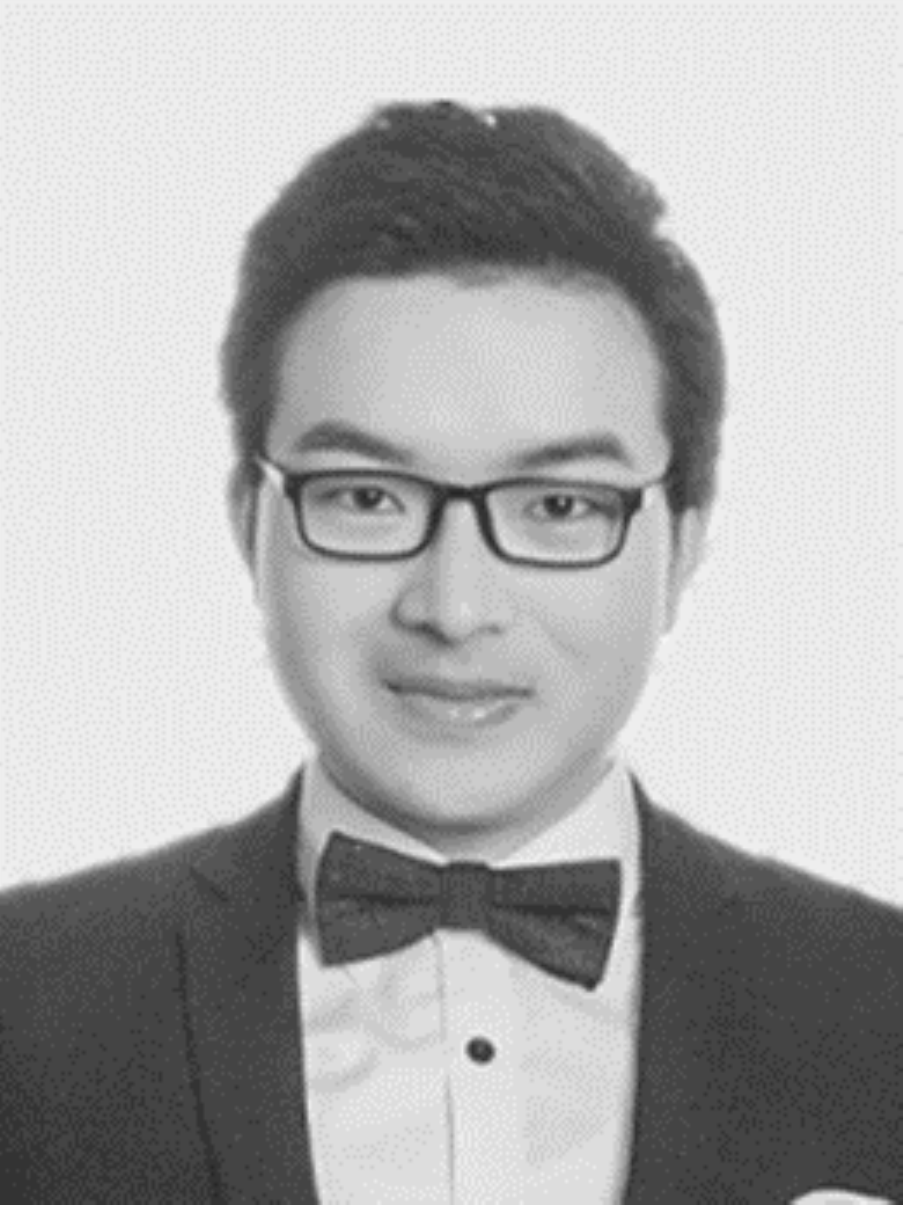}}]{Lizhi Xiong}
	received Ph.D. degree in Communication and Information System from Wuhan University, China in 2016. From 2014 to 2015, he was a Joint-Ph.D. student with Electrical and Computer Engineering, New Jersey University of Technology, New Jersey, USA. He is currently an Associate Professor with School of Computer and Software, Nanjing University of Information Science and Technology, Nanjing, China. His main research interests include privacy-preserving computation, information hiding, and multimedia security.
\end{IEEEbiography}

\begin{IEEEbiography}[{\includegraphics[width=1in,height=1.25in,clip, keepaspectratio]{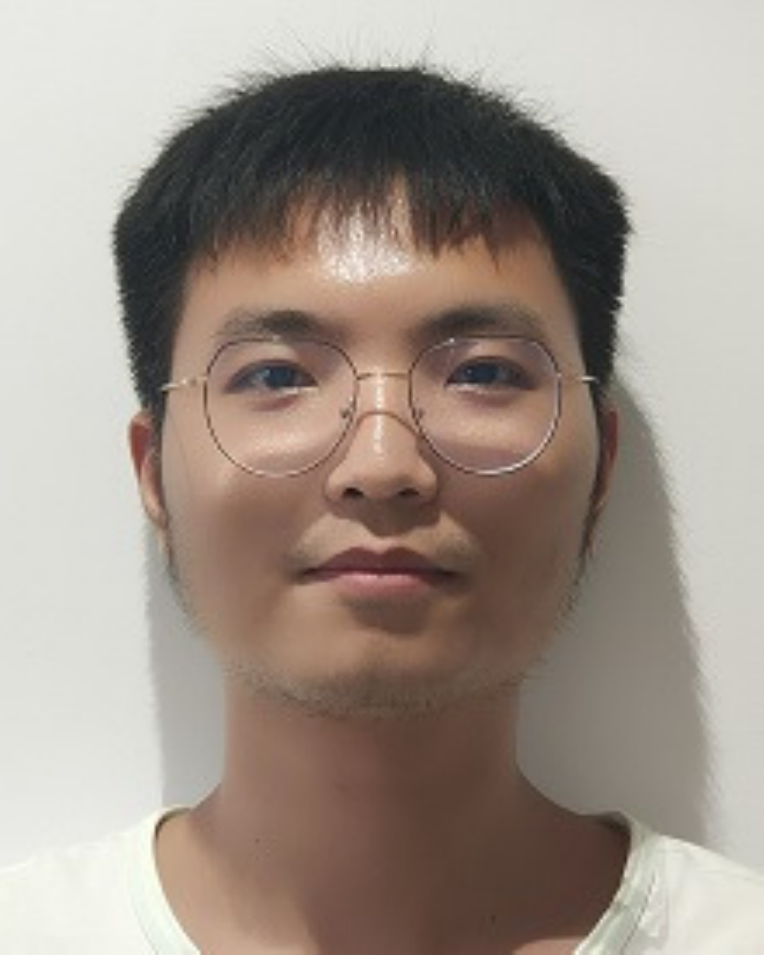}}]{Wenhao Zhou}
	is currently pursing his master degree in the School of Computer and Software, Nanjing University of Information Science and Technology, China. His research interests include secure multiparty computation and privacy-preserving computation.
\end{IEEEbiography}

\begin{IEEEbiography}[{\includegraphics[width=1in,height=1.25in,clip,keepaspectratio]{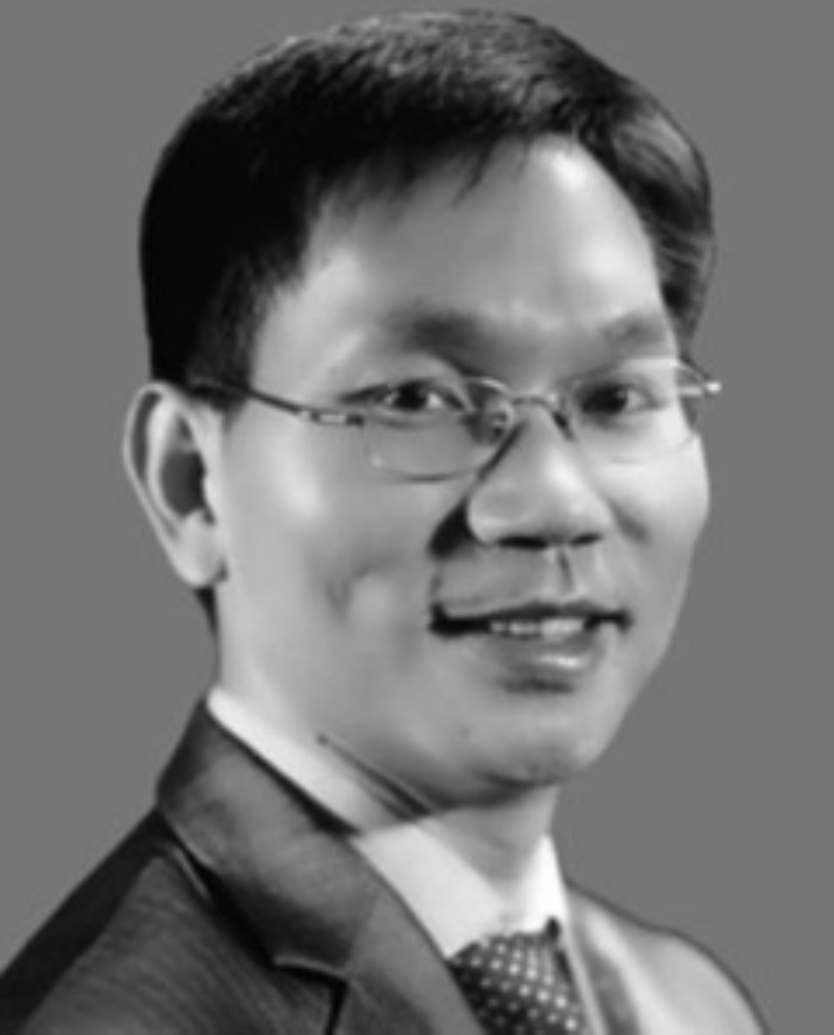}}]{Jian Weng}
	received the Ph.D. degree in computer science and engineering from Shanghai Jiao Tong University, Shanghai, China, in 2008. He is currently a Professor and the Dean with the College of Information Science and Technology, Jinan University,Guangzhou,China. He has authored or coauthored more than 100 papers in cryptography and security conferences and journals, such as CRYPTO, EUROCRYPT, ASIACRYPT, TCC, PKC, TPAMI, TIFS, and TDSC. His research interests include public key cryptography, cloud security, and blockchain. He was the PC Co-Chairs or PC Member for more than 30 international conferences. He also serves as an Associate Editor for the IEEE T RANSACTIONS ON V EHICULART ECHNOLOGY. 
\end{IEEEbiography}

\vfill
	
\end{document}